\documentclass[10pt]{iopart}
\usepackage{iopams}  
\usepackage{graphicx}
\usepackage{color}
\usepackage{hyperref}
\usepackage{comment}
\usepackage[normalem]{ulem}
\newcommand{\pd}{\partial}
\newcommand{\bd}[1]{\boldsymbol{#1}}
\newcommand{\eg}{{e.g., }}

\newcommand{\mc}[1]{\mathcal{#1}}

\newcommand{\avg}[1]{\left\langle{#1}\right\rangle}

\newcommand{\ExB}{{\bd{E}\times \bd{B}}}

\begin{document}

\title[]{Effects of collisional ion orbit loss on tokamak radial electric field and toroidal rotation in an L-mode  plasma}

\author{Hongxuan Zhu, T. Stoltzfus-Dueck, R. Hager, S. Ku, and C.  S. Chang}
\address{Princeton Plasma Physics Laboratory, Princeton, NJ 08543}
\begin{abstract}
Ion orbit loss has been used to model the formation of a strong negative radial electric field  $E_r$  in the tokamak edge, as well as edge momentum transport and toroidal rotation. To quantitatively measure ion orbit loss, an orbit-flux formulation has been developed and numerically applied to the gyrokinetic particle-in-cell code XGC. We study collisional ion orbit loss in an axisymmetric DIII-D L-mode plasma using gyrokinetic ions and drift-kinetic electrons. Numerical simulations, where the plasma density and temperature profiles are maintained
through neutral ionization and heating, show the formation of a quasisteady negative $E_r$ in the edge. We have measured a radially outgoing ion gyrocenter flux due to collisional scattering of ions into the loss orbits, which is balanced by the radially incoming ion gyrocenter flux from confined orbits on the collisional time scale. This suggests that collisional ion orbit loss can shift $E_r$  in the negative direction compared to that in plasmas without orbit loss. It is also found that collisional ion orbit loss can contribute to a radially outgoing (counter-current) toroidal-angular-momentum flux, which is not balanced by the toroidal-angular-momentum flux carried by ions on the confined orbits. Therefore, the edge toroidal rotation shifts in the co-current direction on the collisional time scale.
\end{abstract}
\maketitle
\ioptwocol
\section{Introduction}
Ion orbit loss is considered to have an important impact on radial electric field $E_r$ in the tokamak edge,  where ions leave the confined region and hit the wall due to their finite orbit excursion \cite{Itoh88,Shaing89,Chankin93,Miyamoto96,Connor00,deGrassie09,Chang02}. This effect has been emphasized in diverted tokamaks with a magnetic X point, in which it can significantly depend on the direction of the toroidal magnetic field, and hence the direction of the ion grad-$B$ and curvature drift \cite{Chang02,Shaing02,Ku04,Stacey11,Nishimura20}. Since ions residing in loss orbits do not return to the confined region, ion-orbit loss is often treated as a particle sink in the corresponding loss-orbit portion of phase space. Therefore, in an axisymmetric edge plasma, ions can be continuously scattered into the loss orbits and subsequently leave the confined region. Such collisional ion orbit loss will create a radially outward ion flux, so that $E_r$ will change until the collisional loss-orbit flux is balanced by a radially inward ion flux from the confined orbits \cite{Chang02,Shaing92a,Shaing92b,deGrassie15}. For these reasons, $E_r$ may be different from that in plasmas without orbit loss.

Ion orbit loss has also been used to model the toroidal angular momentum (TAM) transport and toroidal rotation at the outboard of diverted tokamaks \cite{deGrassie09,Muller11a,Solomon11,Muller11b,deGrassie16,Chang08,Pan14,Piper19,Boedo16}. Consider ion orbits at the outboard midplane; ions with parallel velocity in the direction of toroidal plasma current (denoted by ``co-current'') drift radially inwards, while ions with oppositely directed parallel velocity (denoted by ``counter-current'') drift radially outwards (\fref{fig:equil}). Therefore, counter-current ions tend to leave the confined region and subsequently hit the divertor plate or the vessel wall. Assuming the ion distribution function is reduced in the loss-orbit portion of the velocity space, which is mostly counter-current, the total velocity-space distribution function possesses a net co-current momentum. Therefore, ion orbit loss has been linked with the co-current rotation in the tokamak edge. However, such analysis only looks at the velocity space at a given spatial location, and hence cannot describe the global phase-space distribution. For example, many loss orbits are trapped orbits due to their large orbit widths. A trapped orbit consists of both a co-current part and a counter-current part, which are separated by the turning points (banana tips). Consequently, ions residing in this trapped orbit can carry either co-current momentum or counter-current momentum, depending on which flux surface one is looking at. (For example, although  not considered in this paper, loss of fast ions on these trapped orbits can cause significant co-current TAM flux and counter-current toroidal rotation \cite{Helander05NBI,Thyagaraja07}.) Also, such analysis assumes that loss orbits are empty, and hence cannot describe the transport of TAM carried by ions residing in the loss orbits. Quantitative numerical evaluation of the orbit-loss effects on toroidal rotation is therefore desired.

To quantitatively study the effects of ion orbit loss on edge $E_r$ and rotation, an orbit-flux formulation has been developed and numerically applied to the gyrokinetic particle-in-cell code XGC \cite{Stoltzfus20,Stoltzfus21,Zhu22}. This formulation allows quantitative measurements of the loss-orbit contribution to the ion radial gyrocenter particle and momentum flux. In particular, it can distinguish between the various physical mechanisms that contribute to the loss-orbit flux: collisions, turbulent fluctuations, interactions with neutral particles, heating and cooling, and transient effects from time evolution of the plasma. As the first application, this formulation has been used to study collisional ion orbit loss in a DIII-D H-mode plasma \cite{Zhu22}. In reference \cite{Zhu22}, it was found that collisional loss-orbit flux could not significantly change the depth of a simulated H-mode $E_r$ well, which was mostly determined by the steep density pedestal. However, an L-mode plasma has a much shallower density gradient, which may allow the relative orbit-loss contribution to $E_r$ to be more significant.

In this paper, we use the axisymmetric version of XGC (XGCa) to study collisional orbit loss of thermal ions in an axisymmetric DIII-D L-mode plasma with gyrokinetic ions and drift-kinetic electrons \cite{XGC}. Numerical simulations, in which the plasma density and temperature profiles are maintained through neutral ionization and heating, show the formation of a quasisteady negative $E_r$ in the edge. We have measured a radially outgoing ion gyrocenter flux due to collisional scattering of ions into the loss orbits, which is balanced by the radially incoming ion gyrocenter flux from confined orbits on the collisional time scale. This suggests that collisional ion orbit loss can shift $E_r$  in the negative direction, compared to plasmas without orbit loss. It is also found that collisional ion orbit loss can contribute to a radially outgoing (counter-current) TAM flux, which is not balanced by the TAM flux carried by ions on the confined orbits. Therefore, the edge toroidal rotation shifts in the co-current direction on the collisional time scale.

In addition to the main results outlined above, we also studied effects of neutral dynamics and heating on the confined-orbit fluxes, as well as the dependence of ion orbit loss on the plasma density and the direction of the toroidal magnetic field. These preliminary studies yielded interesting results but are not central to the main conclusions of the paper. Therefore, they are presented in appendices.

We note that in a realistic L-mode plasma edge, turbulent particle and momentum transport could be larger than collisional transport \cite{Stoltzfus12,Parra15,Chang17,Ku18,Seo14,Chang21}. In this work, we restrict our attention to purely collisional effects, and leave the effects of turbulent ion orbit loss on edge $E_r$ and toroidal rotation for future study. Magnetic ripple and other 3-dimensional fields can also affect toroidal rotation in tokamaks \cite{Nave10}, but are not included in our simulations, which assume axisymmetric magnetic geometry.

This paper is organized as follows. \Sref{sec:setup} describes the XGCa simulation setup. \Sref{sec:theory} briefly reviews the theory of plasma $E_r$ and and toroidal rotation, as well as our orbit-flux formulation. \Sref{sec:results} presents our simulation results and comparison with theory. Conclusions are given in \Sref{sec:conclusions}. Effects of neutral dynamics and heating on the confined-orbit fluxes are shown in \ref{sec:SOL_flux}. A preliminary study on the density dependence of ion orbit loss is presented in \ref{sec:density}. A comparison with the case when the direction of the toroidal magnetic field is reversed is given in \ref{sec:backward}.

\section{XGCa simulation setup}
\label{sec:setup}
\subsection{Simulation setup}
\label{sec:setup1}
\begin{figure}
    \centering
    \includegraphics[width=1\columnwidth]{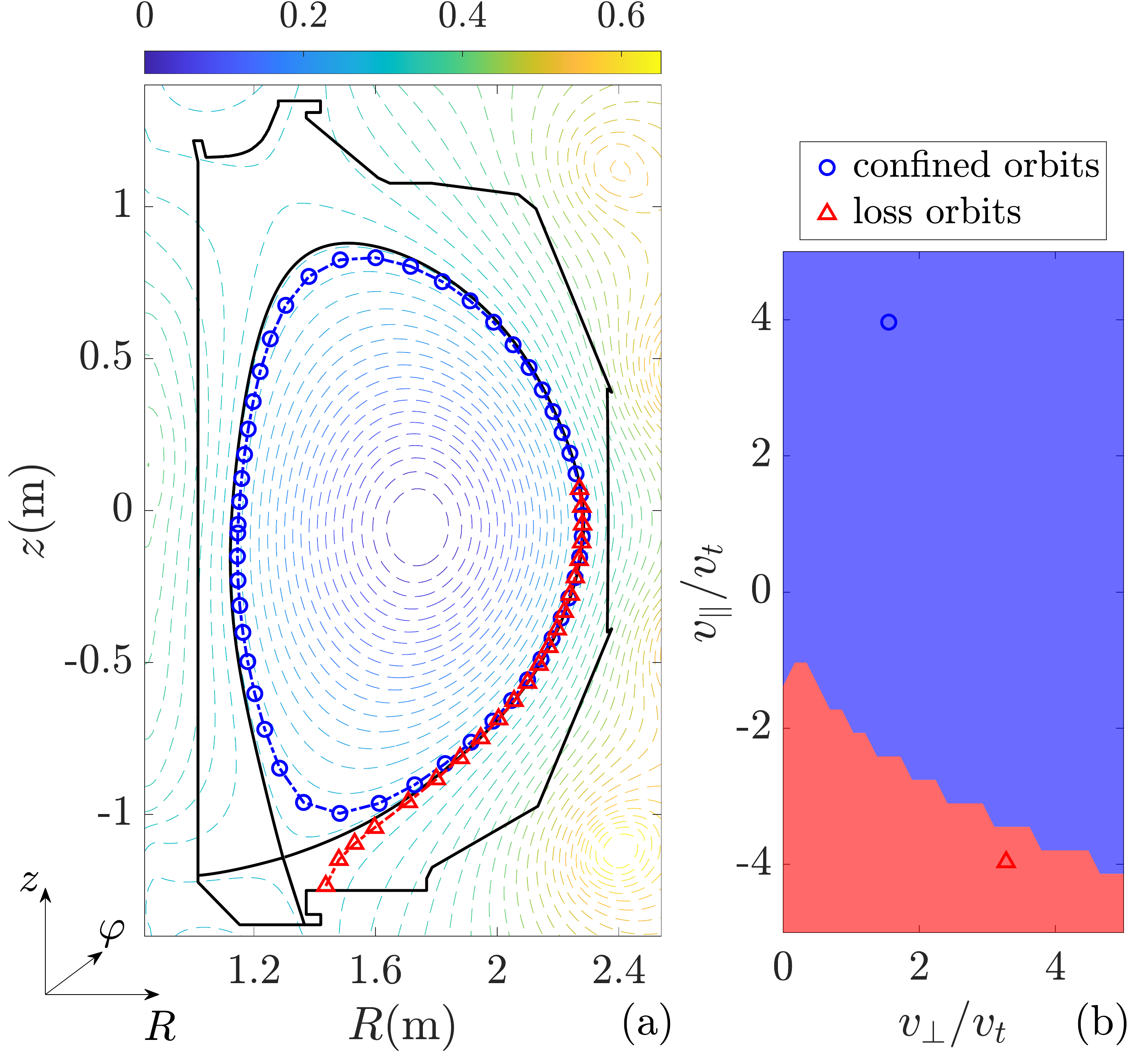}
    \caption{(a) The geometry and equilibrium magnetic field, as well as examples of confined orbits and loss orbits in the edge. The coordinates $(R,\varphi,z)$ form a right-handed coordinate system. The colored dashed-line contours show the poloidal magnetic flux $\psi$. The outer black solid curve shows the vessel wall. The inner black solid curve shows the LCFS containing a magnetic X point, where the corresponding value of $\psi$ is $\psi_X=0.2916{\rm T}\cdot{\rm m}^2$.  The toroidal plasma curent is negative and the corresponding poloidal magnetic field $\bd{B}_\theta=\nabla\psi\times\nabla \varphi$ points in the positive-$\theta$ direction. (We require the poloidal angle $\theta$ to increase counter-clockwise.) The toroidal magnetic field is $\bd{B}_\varphi=I\nabla \varphi$, where $I\approx -3.5{\rm T}\cdot{\rm m}$ is almost constant and hence not shown. The blue circles show a co-current confined orbit passing through the outboard midplane of the $\psi_n=0.99$ flux surface. The red triangles show a counter-current loss orbit passing through the same point. (b) Velocity-space distribution of confined orbits (blue color) and loss orbits (red color) at the outboard midplane of the $\psi_n=0.99$ flux surface. Here, $v_\perp\doteq\sqrt{2\mu B/m_i}$, $v_\parallel\doteq p_\parallel/m_i$, and the local ion thermal velocity is $v_t$. The blue circle and the red triangle correspond to the two orbits shown in (a). More examples of loss orbits can be found, \eg in reference \cite{Ku04}.}
    \label{fig:equil}
\end{figure}
We use electrostatic XGCa simulations to study an axisymmetric L-mode plasma in DIII-D geometry  from the magnetic axis to the wall (\fref{fig:equil}). The simulation setup is similar to that in \cite{Zhu22}, but with differences emphasized below. The code uses cylindrical coordinates $\bd{R}\doteq(R,\varphi,z)$ to describe the realistic toroidal geometry containing an X point, where $\doteq$ means definitions. The simulation domain includes the confined region, the last closed flux surface (LCFS), the scrape-off layer (SOL), and the wall. The equilibrium magnetic field is given by $\bd{B}=I(\psi)\nabla\varphi+\nabla\psi\times\nabla\varphi$, where $\psi$ is the poloidal magnetic flux and $I=RB_\varphi$ is a flux function. The toroidal magnetic field $B_\varphi$ is negative, so that the resulting ion curvature drift points in the negative-$z$ direction.  The toroidal plasma current is negative, so that the corresponding poloidal magnetic field $\bd{B}_\theta$ points in the counter-clockwise direction. 
The toroidal coordinates $(\psi,\varphi,\theta)$ are not directly used in the code, but can be used to describe simulation results. Here, we can define the poloidal angle as $\theta=\tan^{-1}(z-z_a)/(R-R_a)$ where $(R_a,z_a)$ is the location of the magnetic axis. We also require $\theta\in[0,2\pi)$ to increase counter-clockwise.
In the following, ``radial'' refers to the direction perpendicular to flux surfaces labeled by $\psi$, so the radial electric field is defined as $E_r\doteq\bd{E}\cdot\nabla\psi/|\nabla\psi|$. Since $E_r$ varies in the poloidal direction, we look at its value at the outside midplane, $\theta=0$. Also, ``poloidal'' refers to the direction of $\bd{e}_\theta\doteq\pd\bd{R}/\pd\theta$, which is tangent to flux surfaces. For example, the poloidal magnetic field is $B_\theta\doteq\bd{B}\cdot\bd{e}_\theta/|\bd{e}_\theta|$, which is positive within our simulations. The code utilizes unstructured triangular meshes, with most of the mesh nodes aligned with magnetic field lines \cite{Zhang16}. The radial grid size is chosen based on the inverse of the local ion radial density gradient, while the poloidal grid size is chosen based on the local ion gyroradius. Near the LCFS, the  radial grid size is $\Delta\psi_n\approx 0.002$, which is $\Delta R\approx 0.8{\rm mm}$ at the outboard midplane, and the poloidal grid size is $\Delta l_\theta\approx 5{\rm mm}$.  Here, the normalized flux is defined as $\psi_n\doteq(\psi-\psi_{\rm a})/(\psi_X-\psi_{\rm a})$, where $\psi_{\rm a}$ and $\psi_X$ are the value of $\psi$ at the magnetic axis and at the LCFS, respectively.

\begin{figure}
    \centering
    \includegraphics[width=1\columnwidth]{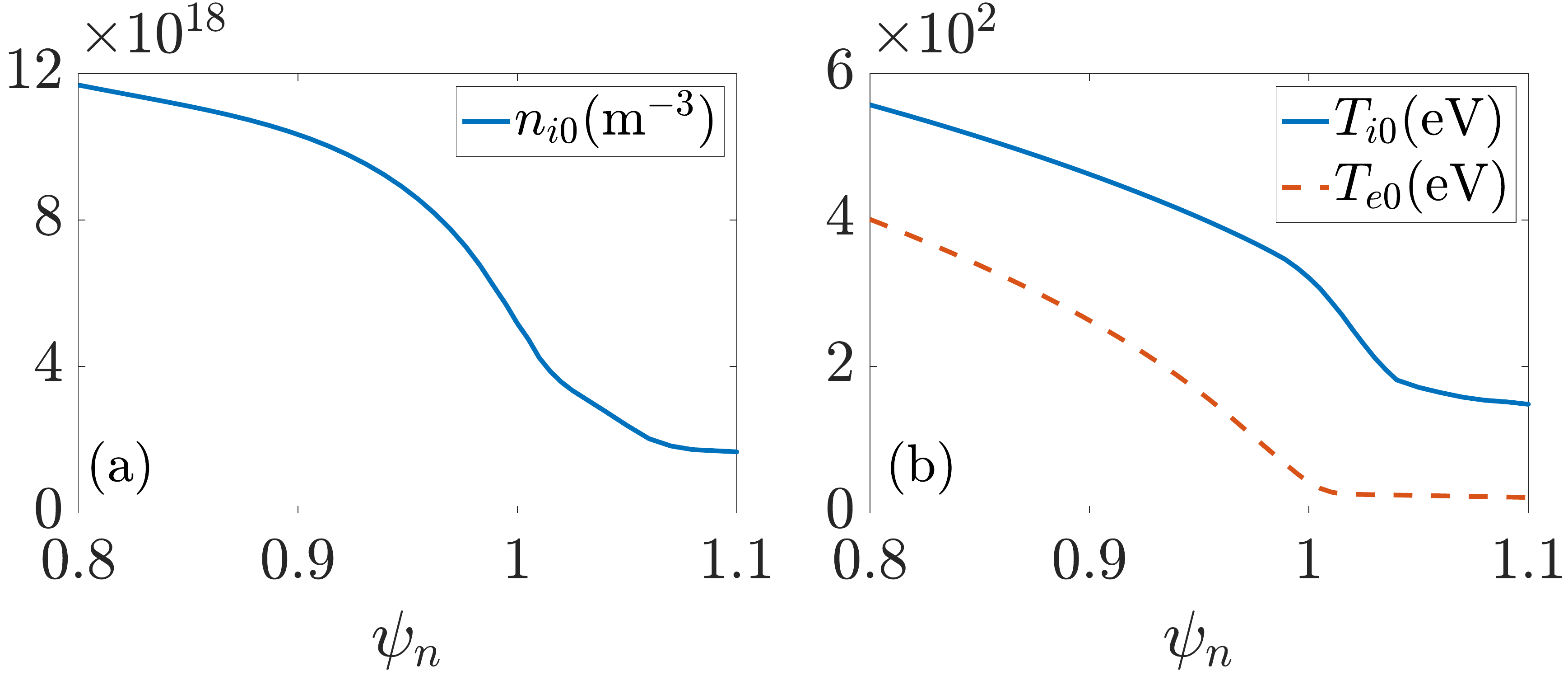}
    \caption{The equilibrium plasma density (a) and temperature (b) versus the normalized flux  $\psi_n\doteq(\psi-\psi_{\rm a})/(\psi_X-\psi_{\rm a})$ in the edge. Here, $\psi_{\rm a}$ and $\psi_X$ are the value of $\psi$ at the magnetic axis and the X point, respectively. The electron density $n_{e0}$ equals the ion density $n_{i0}$ due to quasineutrality.}
    \label{fig:Lprofile}
\end{figure}

We simulate gyrokinetic deuterium ions and drift-kinetic electrons. We retain only a single species of thermal ions, so effects from impurities and fast ions are not considered. Their equilibrium density $n_{s0}(\psi)$ and temperature profiles $T_{s0}(\psi)$ are flux functions (figure \ref{fig:Lprofile}), and are from shot 161218 \cite{Yan17,Wang18,Schmitz22}. Species indices are $s=i$ for ions and $s=e$ for electrons. The coordinates  are position $\bd{R}\doteq(R,\varphi,z)$, magnetic moment $\mu$, and parallel momentum $p_\parallel$. The characteristics are governed by equations given in \cite{Chang04}, which are mathematically equivalent to the following:  
\begin{eqnarray}
\label{XGC_Rdot}
B_\parallel^*\dot{\bd{R}}=(Z_se)^{-1}\hat{\bd{b}}\times\nabla H+v_\parallel \bd{B}^*,\\
\label{XGC_pdot}
B_\parallel^* \dot{p}_\parallel=-\bd{B}^*\cdot\nabla H,
\end{eqnarray}
and $\dot{\mu}=0$. Here, the overhead dot denotes the time derivative, $\hat{\bd{b}}\doteq\bd{B}/B$, $\bd{B}^* \doteq\bd{B}+\nabla\times(p_\parallel\hat{\bd{b}}/Z_s e)$, $B_\parallel^*\doteq\hat{\bd{b}}\cdot\bd{B}^*$,  $H=p_\parallel^2/2m_s+\mu B+Z_se \hat{J}_0{\Phi}$ is the Hamiltonian, $v_\parallel\doteq\pd_{p_\parallel}H$ is the parallel velocity, $e$ is the elementary charge, $Z_s$ is the charge number, $\hat{J}_0$ is the gyroaveraging operator ($\hat{J}_0=1$ for drift-kinetic electrons), and $\Phi$ is the electrostatic potential. 

Using the total-f simulation method \cite{Ku16}, the code calculates the phase-space distribution functions $F_s(\bd{R},\mu,p_\parallel,t)$ of ion and electron gyrocenters. The distribution functions are chosen to be Maxwellian at $t=0$. Then, ion and electron markers' weights evolve such that $F_s$ advances in time according to 
\begin{equation}
\label{XGC_vlasov}
d_tF_s\doteq\pd_t F_s+\dot{\bd{R}}\cdot\nabla F_s+\dot{p}_\parallel \pd_{p_\parallel}F_s=C_s+S_s+N_s.
\end{equation}
Here, $C_s$ is a fully nonlinear multi-species Fokker–-Planck–-Landau collision operator \cite{Yoon14,Hager16}, $S_s$ describes external heating, and $N_s$ describes neutral ionization and charge exchange. The use of $S_s$ and $N_s$ will be further discussed below. The logical-sheath boundary condition is used at the wall, where all ions are absorbed, and electrons are absorbed if their energy is higher than the sheath potential energy $e\Phi_{\rm sh}$; otherwise electrons are reflected \cite{Ku18,Parker93}.

The electrostatic potential $\Phi$ is calculated from the gyrokinetic Poisson equation. With $Z_i=1$ and $Z_e=-1$, it is written as
\begin{equation}
\label{XGC_poisson}
\nabla_\perp\cdot\left(\frac{n_{i0}m_i}{eB^2}\nabla_\perp\Phi\right)=-(\delta \bar{n}_i-\delta n_e),
\end{equation}
where $\nabla_\perp$ denotes gradient perpendicular to $\bd{B}$. Here, $\delta\bar{n}_i\doteq\int d\mc{W}\hat{J}_0 F_i -n_{i0}$, $\delta{n}_e\doteq\int d\mc{W}F_e -n_{e0}$, and $\int d\mc{W}\doteq(2\pi/m_i^2)\int d\mu\,dp_\parallel B_\parallel^*$ denotes velocity-space integration. In our simulations, only the radial component of the electric field $\bd{E}=-\nabla\Phi$ is used, while the poloidal component is neglected for simplicity. In other words, $\Phi$ is treated as a flux function in our simulations. In the SOL, $\Phi$ is set to be equal to the sheath potential $\Phi_{\rm sh}$, which is initialized with a theoretical value assuming Maxwellian ions and electrons \cite{Stangeby00book}:
\begin{equation}
\label{shpot}
    e\Phi_{\rm sh}(t=0)=-\frac{T_{e0}}{2}\ln\left[\frac{2\pi m_e}{m_i}(1+\frac{T_{i0}}{T_{e0}})\right].
\end{equation}
At $t>0$, $\Phi_{\rm sh}$ is adjusted such that the electron loss rate matches the ion loss rate at the wall. However, $\Phi$ is always set to $\Phi_{\rm sh}(t=0)$ in the SOL in our simulations. This helps avoid fluctuations in $E_r$ in the SOL and thus make the simulations stable. Parallel variation of $\Phi$ is also ignored in the SOL to achieve better numerical stability.  In experiments, parallel electric fields arise to confine electrons in the SOL, and the corresponding parallel variation of $\Phi$ is on the order of $T_e/e$. Since we focus on $E_r$ and toroidal rotation in the confined region, we do not expect to accurately model electric fields in the SOL. Nevertheless, by assuming $\Phi=\Phi_{\rm sh}$, the SOL $E_r$ is a few times $-\pd_r T_e/e$, within the approximate range of $E_r$ that is expected in an experimental SOL.
\subsection{Maintain the L-mode profile using neutral ionization and heating}
\label{sec:setup2}
Although effects from collisions are the central focus of this study, neutral ionization and heating have been included in order to maintain the L-mode profile at the edge. After the simulation begins, ions on loss orbits are quickly lost to the wall. Without a source of particles, such collisionless orbit loss will result in a quick drop of the ion density inside the LCFS, which results in formation of a density pedestal and large negative $E_r$ there \cite{Chang04}. To prevent the density loss, a Monte Carlo neutral model is included \cite{Ku18} and is described by $N_s$ in \eref{XGC_vlasov}. In our simulations, all ions hitting the wall return to the plasma as neutral particles; namely, the neutral recycling coefficient is set to one. These neutral particles are low in energy, and experience random ionization and charge exchange while moving from the wall to the plasma. Neutral particles ionize due to impact with electrons; after ionization, the impact electrons lose some energy, while the neutral particles become pairs of low-energy ions and electrons. For the charge exchange, neutrals transfer their electrons to thermal ions, so thermal ions are replaced by low-energy ions. 

Simulation results show that neutral particles can penetrate deeply into the confined region if we use physical ionization and charge-exchange rates, causing too much density buildup in the confined region. Therefore, to maintain the density profile, we manually increase the ionization and charge-exchange rates so that most neutral particles ionize in the SOL and do not penetrate deeply into the confined region. To be clear, we do not intend to study neutral physics here; instead, neutrals are merely used as a particle source to maintain the density profile.

Since neutral ionization and charge exchange create low-energy ions and electrons, the plasma temperature quickly drops in the edge. To maintain the temperature profile, ion and electron heating is applied in the edge, which is described by $S_s$ in \eref{XGC_vlasov}. The heating is only applied in the SOL region, and the power is dynamically adjusted based on the ion and electron heat flux to the wall. In our simulations, $S_s$ does not generate net plasma density or momentum.

\begin{figure}
    \centering
    \includegraphics[width=1\columnwidth]{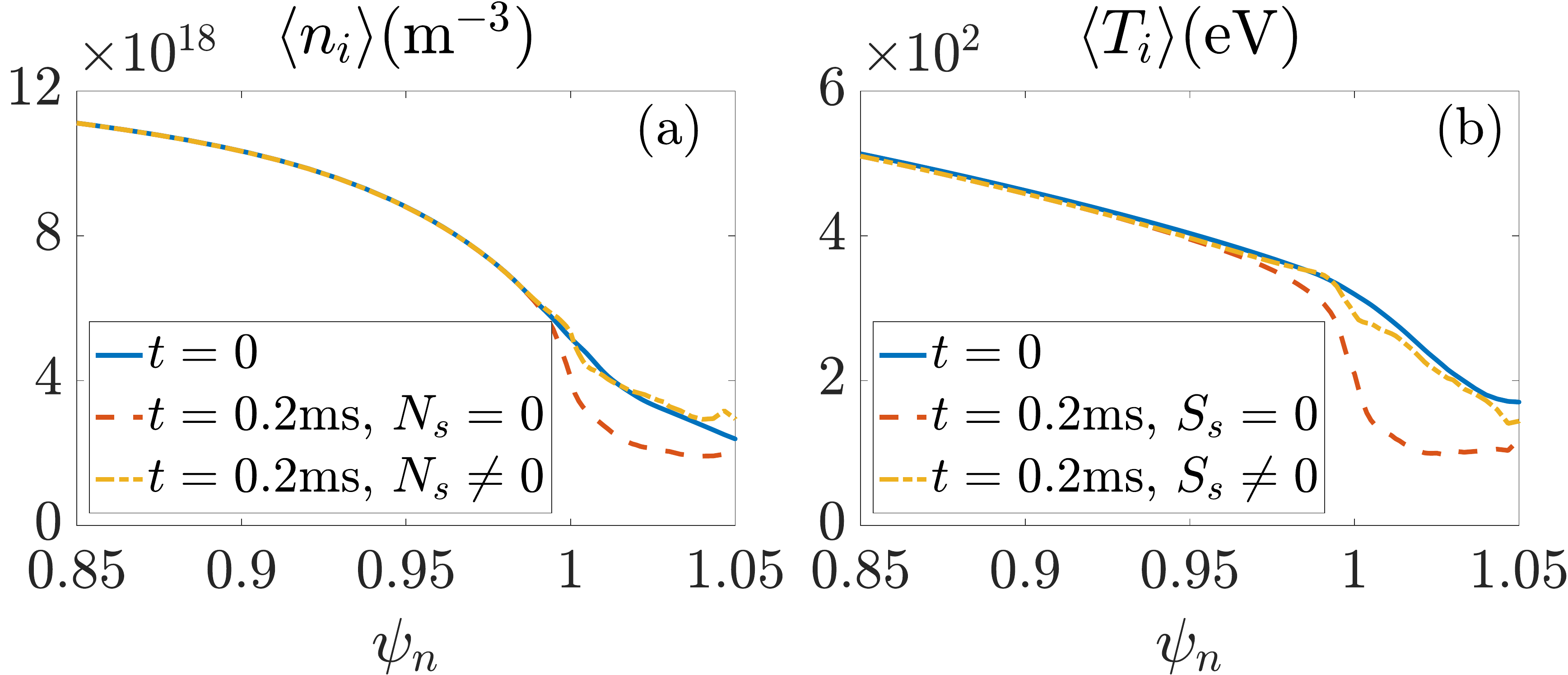}
    \caption{(a) The flux-surface averaged ion density versus $\psi_n$. Blue solid line: initial value. Red dashed line: the value at $t=0.2$ms without neutral ionization. Yellow dot-dashed line: the value at $t=0.2$ms with neutral ionization. (b) The flux-surface averaged ion temperature versus $\psi_n$. Blue solid line: initial value. Red dashed line: the value at $t=0.2$ms with neutral ionization but without heating. Yellow dot-dashed line: the value at $t=0.2$ms with neutral ionization and heating.}
    \label{fig:Lprofile2}
\end{figure}
\Fref{fig:Lprofile2} demonstrates that the edge $n_i$ and $T_i$ profiles are maintained through neutral ionization and heating. Due to the resulting shallow density gradient, the negative $E_r$ well remains shallow, which allows us to evaluate ion orbit loss in an L-mode edge. We note that it is difficult to tailor  $N_s$ and $S_s$ in order to maintain the profiles exactly. Instead, they are only maintained approximately and still evolve somewhat over time. There is also significant poloidal asymmetry in $n_i$ and $T_i$ in the edge (\sref{sec:flow}), so that the 2-dimensional plasma profiles are always different from initial conditions. Additionally, since $T_i>T_e$, ions transfer energy to electrons via collisions, so that $T_i$ decreases while $T_e$ increases, even in the confined region away from the edge. Therefore, we do not expect the simulation to reach a truly steady state using the current simulation setup.
\section{Theoretical background}
\label{sec:theory}
\subsection{The plasma radial electric field and toroidal rotation}
\label{sec:neo}
The neoclassical theory of $E_r$ has been well established in the plasma core where $\delta\doteq\rho_{i\theta}/L\ll 1$ \cite{Hinton73,Hazeltine74,Hirshman78,Hirshman81,Helander05book,Dorf13}. Here, $\rho_{i\theta}$ is the ions' poloidal gyroradius and $L$ is the radial scale length of ions' equilibrium profile. Transient behaviors, such as the geodesic-acoustic mode, quickly damp due to  ion Landau damping and collisions. Neoclassical theory solves the ion drift-kinetic equation assuming slow time evolution, effectively considering the system after the decay of transients. To the lowest order in $\delta$, the ion fluid velocity $\bd{u}_i$ can be written as
\begin{equation}
\label{fluid_flow}
    \bd{u}_i=u_{i\parallel}\hat{\bd{b}}+\frac{\hat{\bd{b}}\times\nabla\Phi}{B}+\frac{\hat{\bd{b}}\times\nabla p_i}{Z_ien_iB},
\end{equation}
where $u_{i\parallel}$ is the ion parallel velocity and $p_i\doteq n_iT_i$ is the ion pressure. Again to lowest order in $\delta$, the ion distribution function is Maxwellian, and both $\Phi$ and $p_i$ are flux functions. Then, neoclassical theory neglects radial drifts and assumes an incompressible particle flux, which restricts $\bd{u}_i$ to the following form:
\begin{equation}
\label{neo_flow}
    \bd{u}_i=\omega_i R\hat{\bd{\varphi}}+K\bd{B}/n_i,
\end{equation}
where $\omega_i=\pd_\psi\Phi+(\pd_\psi p_{i})/(en_{i})$ describes toroidal rotation, and $K=K(\psi)$ is a flux function that describes poloidal rotation. According to neoclassical theory, $K/n_iI\avg{B^{-2}}=-k \pd_\psi T_i/e$, where the coefficient $k$ is the neoclassical constant of proportionality between ion poloidal rotation and the ion temperature gradient, and $k$ depends on the collisionality. The flow velocity \eref{neo_flow} is incompressible, $\nabla\cdot (n_i\bd{u}_i)=0$, and produces the correct diamagnetic and $\ExB$ drift velocity in the direction perpendicular to $\bd{B}$.  In the core, the TAM density $\mc{L}_\varphi\approx\avg{m_in_{i}R u_{i\parallel}B_\varphi/B}$ is conserved for each flux surface on the collisional time scale, where $\avg{\dots}$ denotes flux-surface averaging. Therefore, we have a unique solution of $E_r$ in steady state:
\begin{equation}
\label{Er_neo}
    \pd_\psi\Phi=-\frac{\pd_\psi p_{i}}{en_{i}}-\frac{K}{n_{i}I\avg{B^{-2}}}+\frac{\mc{L}_\varphi}{m_in_{i} I^2\avg{B^{-2}}}.
\end{equation}
The TAM density $\mc{L}_\varphi$ is conserved and is determined by the initial condition. For our simulations, the initially Maxwellian distribution produces no parallel flows or electric fields, so that $\mc{L}_\varphi$ is of order $(B_\theta^2/B^2)$, which is usually quite small and can be neglected in the core.

Note that the fluid theory calculates velocity of actual ion particles instead of ion gyrocenters. The difference between the ion particle flux from \eref{fluid_flow} and the ion gyrocenter flux from \eref{XGC_Rdot} is known as the magnetization flux \cite{Chankin97,Brizard07}, which is incompressible and hence does not affect the above analysis. Therefore, the ion fluid velocity \eref{fluid_flow} can be readily used to analyze the results from gyrokinetic simulations.

The above standard neoclassical theory no longer works in the edge, where the assumption $\delta\ll 1$ usually breaks down. The loss orbits connect to the wall, and hence the corresponding distribution function cannot be approximately Maxwellian. There is also a significant radial TAM flux at the edge, causing the local TAM to evolve over time. A comparison between theory and simulation results will be given in sections~\ref{sec:Er} and \ref{sec:flow}.

Since toroidal rotation is important in the edge, we also look at the edge TAM density and flux. Within the gyrokinetic formulation, the total TAM density of ion and electron gyrocenters can be written as 
\begin{equation}
    \mc{L}_\varphi=\sum_s \mc{L}_{s\varphi}+\mc{L}_E,
\end{equation}
 where $\mc{L}_{s\varphi}$ is the parallel-flow portion of the TAM density,
\begin{equation}
    \mc{L}_{s\varphi}\doteq \avg{\int d\mc{W}F_sp_\parallel R B_\varphi/B},
\end{equation}
and $\mc{L}_E$ is the $\ExB$ portion of the TAM density,
\begin{equation}
\label{TAM_LE}
    \mc{L}_E\doteq-\avg{\bd{P}\cdot\nabla\psi},
\end{equation}
with the polarization $\bd{P}$ being the solution of
\begin{equation}
    \nabla\cdot\bd{P}\doteq\sum_sZ_se\int d\mc{W}F_s.
\end{equation}
The total gyrocenter radial TAM flux is
\begin{equation}
    \Pi_r=\sum_s\Pi_{sr}\doteq\sum_s\int d\mc{W} F_s p_\parallel R(B_\varphi/B)\dot{\bd{R}}\cdot d\bd{S},
\end{equation}
where $d\bd{S}$ is the surface element of the given flux surface. Then, the gyrokinetic equation \eref{XGC_vlasov} results in the following exact TAM conservation relation for axisymmetric plasmas \cite{Scott10,Stoltzfus17,Abiteboul11}:
\begin{equation}
\label{TAM_conservation}
    \pd_t\mc{L}_\varphi=-\pd_V\Pi_{r},
\end{equation}
where $V(\psi)$ is the volume inside the flux surface. \Eref{TAM_conservation} applies to the confined region with closed flux surfaces, including edge flux surfaces where loss orbits are present.
\subsection{Decomposition of ion gyrocenter particle and momentum flux}
\label{sec:orbit_flux}
In our axisymmetric simulations, there are no turbulent electric fields and thus no turbulent radial electron density flux. The electron radial density flux due to orbit excursion is much smaller than that of ions. Collisions will create nonzero ion and electron radial fluxes, which are ambipolar due to momentum conservation and hence do not directly contribute to $E_r$. Neutral ionization does not change $E_r$ directly either, as it creates equal number of ions and electrons. Therefore, in our axisymmetric simulations, the edge $E_r$ is mostly determined by ion dynamics. The orbit-flux formulation \cite{Stoltzfus20,Stoltzfus21} determines separate contributions to the ion gyrocenter radial flux from collisions, heating, neutral ionization and charge exchange, and time evolution of the plasma. Specifically, if one solves the orbit characteristics \eref{XGC_Rdot} and \eref{XGC_pdot} using the value of $H$ at fixed $t$, one can obtain a set of instantaneous orbits $(\bd{R}(\tau),p_\parallel(\tau))$, where $\tau$ is a timelike variable that parameterizes the orbits. Since $H$ is axisymmetric, the canonical toroidal angular momentum $\mc{P}_\varphi\doteq Z_ie\psi+p_\parallel\hat{\bd{b}}\cdot R^2\nabla\varphi$ is conserved, and hence these orbits can be labeled by $(\mu,\mc{P}_\varphi,H)$. Through a coordinate transformation, one can show that the ion radial gyrocenter flux $\Gamma_{ir}$ across a given flux surface can be written as
\begin{equation}
\label{formulation_flux}
\eqalign{
\Gamma_{ir}=\int d\bd{S}\cdot\int d\mc{W} F_i{\dot{\bd{R}}}=\frac{2\pi}{Z_iem_i^2}\times\\
\int_0^{\infty} d\mu\int_{-\infty}^{\infty} d\mc{P}_\varphi\int_{H_{\min}}^{H_{\max}} d{H}\oint d\varphi\int_0^{\tau_{\rm orb}} d\tau (d_\tau F_i).
}
\end{equation}
Here, ${d}_\tau\doteq{\dot{\bd{R}}}\cdot\nabla+{\dot{p}}_\parallel\pd_{p_\parallel}$ is the derivative along instantaneous orbits at fixed time $t$. From \eref{XGC_vlasov}, the ion distribution function follows
\begin{equation}
\label{formulation_vlasov}
{d}_\tau F_i=C_i+S_i+N_i-\pd_t F_i.
\end{equation}
Therefore, combining \eref{formulation_flux} and \eref{formulation_vlasov}, the ion radial flux is decomposed into the contribution from each term:
\begin{equation}
\label{formulation_terms}
\Gamma_{ir}=\Gamma_{\rm col}+\Gamma_{\rm heat}+\Gamma_{\rm neut}+\Gamma_t.
\end{equation}
Here, $\Gamma_{\rm col}$ is from  $C_i$,  $\Gamma_{\rm heat}$ is from  $S_i$, $\Gamma_{\rm neut}$ is from $N_i$, and $\Gamma_t$ is from $-\pd_t F_i$. 

\Eref{formulation_terms} describes how the radial ion gyrocenter flux is sustained through various physical mechanisms. For example, if collisions can continuously scatter ions into the orbits in steady state, then there will be more ions leaving the flux surface than those entering, giving a positive contribution to $\Gamma_{ir}$ from $\Gamma_{\rm col}$.  Similar interpretations can be applied to $\Gamma_{\rm heat}$ and $\Gamma_{\rm neut}$. Meanwhile,  $\Gamma_t$ is the time variation of the number of ions residing in the orbits and only describes transient behavior of the plasma. One expects $\Gamma_t$ to vanish in steady states, but that is not achieved in our simulations of an axisymmetric plasma, and hence this term cannot be ignored in \eref{formulation_terms}.

In the edge, one can further decompose orbits into loss orbits, which connect to the wall, and confined orbits, which form closed loops and do not connect to the wall (\fref{fig:equil}). Then, each term in \eref{formulation_terms} can be decomposed into the loss-orbit contribution and the confined-orbit contribution, \eg
\begin{equation}
   \Gamma_{\rm col}= \Gamma_{\rm col}^{\rm loss}+\Gamma_{\rm col}^{\rm conf}.
\end{equation}
Here, $\Gamma_{\rm col}^{\rm loss}$ (or $\Gamma_{\rm col}^{\rm conf}$) is the radial flux due to collisional scattering of ions into the loss orbits (or confined orbits).  Similar interpretations can be applied to $\Gamma_{\rm heat}$ and $\Gamma_{\rm neut}$. Meanwhile, $\Gamma_t^{\rm loss}$ (or $\Gamma_t^{\rm conf}$) describes the time variation of the number of ions residing in loss orbits (or confined orbits).

The above formulation can also be used to calculate radial TAM flux of ion gyrocenters. The TAM associated with the parallel motion of ion gyrocenters are $p_\parallel\hat{\bd{b}}\cdot R^2\nabla\varphi=\mc{P}_\varphi - Z_ie\psi$. Therefore, the ion gyrocenter TAM flux $\Pi_{ir}$ across a given flux surface can be written as
\begin{equation}
\label{formulation_flux2}
\eqalign{
\Pi_{ir}=\int d\bd{S}\cdot\int d\mc{W} (\mc{P}_\varphi-Z_ie\psi)F_i{\dot{\bd{R}}}=\frac{2\pi}{Z_iem_i^2}\times\\
\int d\mu\,d\mc{P}_\varphi\,dH\,d\varphi\,d\tau(\mc{P}_\varphi-Z_ie\psi) (d_\tau F_i)\\
=\Pi_{\rm col}+\Pi_{\rm heat}+\Pi_{\rm neut}+\Pi_t.
}
\end{equation}
Similar to the particle flux, each term in the TAM flux can be further decomposed into contributions from the loss orbits, and contributions from the confined orbits.

Finally, note that the orbit integration in \eref{formulation_flux} is along the part of the orbit inside the given flux surface. Since confined orbits form closed loops, we can also integrate along the part outside the flux surface. For any confined orbit, we have
\begin{equation}
    \int_{\rm inside} d\tau (d_\tau F_i)=-\int_{\rm outside} d\tau (d_\tau F_i),
\end{equation}
and the following relation for the confined-orbit fluxes:
\begin{equation}
\label{formulation_outside}
\eqalign{
    \left(\Gamma_{\rm col}^{\rm conf}+\Gamma_{\rm heat}^{\rm conf}+\Gamma_{\rm neut}^{\rm conf}+\Gamma_t^{\rm conf}\right)|_{\rm inside}\\
    =-\left(\Gamma_{\rm col}^{\rm conf}+\Gamma_{\rm heat}^{\rm conf}+\Gamma_{\rm neut}^{\rm conf}+\Gamma_t^{\rm conf}\right)|_{\rm outside}.
    }
\end{equation}
\Eref{formulation_outside} does not apply to the loss-orbit fluxes, though, since loss orbits connect to the wall instead of coming back to the confined region.
\section{Simulation results}
\label{sec:results}
\begin{figure*}
    \centering
    \includegraphics[width=1\textwidth]{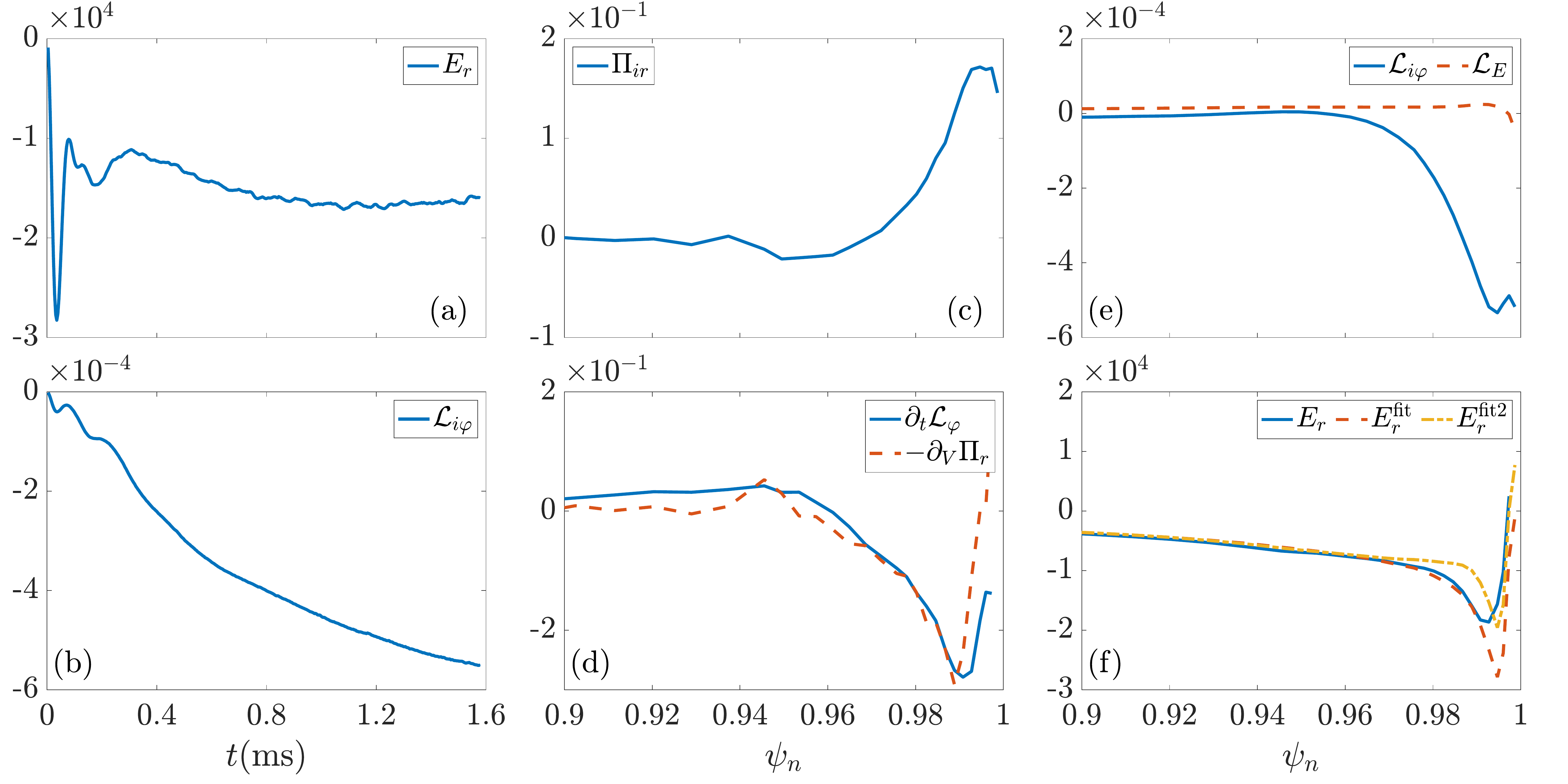}
    \caption{(a) The radial electric field (in units V/m) versus $t$ at the outboard midplane of the $\psi_n=0.99$ flux surface. (b) The ion TAM density (in units $\mathrm{kg/(m\cdot s)}$) at the same flux surface. (c) The radial ion TAM flux (in units $\mathrm{kg\cdot m^2/s^2}$) versus $\psi_n$ at $t=0.8$ms. (d) Comparison between $\pd_t\mc{L}_\varphi$ and $-\pd_V\Pi_r$ (in units $\mathrm{kg/(m\cdot s^2)}$) versus $\psi_n$ at $t=0.8$ms. The data is averaged over a time window $\Delta t=0.08$ms. (e) Comparison of $\mc{L}_{i\varphi}$ and $\mc{L}_E$ at the edge at $t=0.8$ms. (f) Comparison of $E_r$ with the fitting formulae \eref{Er_fit} and \eref{Er_fit2}.}
    \label{fig:Er1}
\end{figure*}

In this section, we present simulation results of the L-mode plasma edge. We focus on edge $E_r$ and toroidal rotation, and compare them with the theory outlined above. It is found that the behaviors of $E_r$ and toroidal rotation cannot be predicted by neoclassical theory. Numerical orbit-flux results showed that for the gyrocenter particle flux, the loss-orbit contribution is nearly balanced by the confined-orbit contribution, leaving a small positive ambipolar value in the sum. This suggests that the steady-state $E_r$ can be different from that in a plasma without orbit loss. Meanwhile, for the gyrocenter TAM flux, the loss-orbit contribution is not balanced by the confined-orbit contribution, leaving a radially outward counter-current momentum flux in the sum.
\subsection{Edge $E_r$ and toroidal rotation}
\label{sec:Er}
We ran the simulation for 1.6ms, which is approximately one ion--ion collision time $\nu_i^{-1}$ for our edge parameters $n_i=6\times 10^{18}{\rm m}^{-3}$ and $T_i=300$eV. At the beginning of the simulation, geodesic-acoustic-mode oscillations are excited, but are quickly damped due to collisions and ion Landau damping, after which $E_r$ in the core relaxes to its neoclassical solution \eref{Er_neo}. In the edge, however, the plasma is not in neoclassical equilibrium, due to open field lines, ion orbit loss, neutral ionization, and heating, so we do not expect a simple relaxation to neoclassical $E_r$. Nevertheless, a negative edge $E_r$ quickly forms due to the ion radial pressure gradient, as expected from neoclassical theory. \Fref{fig:Er1}(a) shows an example of edge $E_r$ versus $t$ at the $\psi_n=0.99$ flux surface. We choose this flux surface because we have set $\Phi=\Phi_{\rm sh}$ in the SOL (\sref{sec:setup1}), which decreases radially and thus enforces a positive $E_r$ there. This means that $E_r$ transitions from being negative to positive across the LCFS, introducing some uncertainty in its value around $\psi_n\approx 1$. (Turbulent viscosity may be needed to model the transition of $E_r$ near the LCFS \cite{Rozhansky06,Staebler15}.) At the $\psi_n=0.99$ flux surface, $E_r$ reaches a relatively steady value at $t>0.4$ms, and only evolves slowly thereafter. 

While $E_r$ does not change much, there is a toroidal-rotation acceleration at the edge. \Fref{fig:Er1}(b) shows the ion gyrocenter parallel-flow TAM density $\mc{L}_{i\varphi}$ at the $\psi_n=0.99$ flux surface, which shifts in the negative direction. This is the same direction as the equilibrium toroidal current that produces $B_\theta$; namely, the toroidal-rotation acceleration is co-current. This co-current toroidal rotation cannot be described by standard neoclassical theory, which predicts small radial TAM flux and hence small toroidal rotation. Note that the acceleration of $\mc{L}_{i\varphi}$ gradually slows down over time, which will be further discussed in \sref{sec:results_flux}.

The co-current toroidal rotation acceleration is due to a positive (counter-current) ion gyrocenter TAM flux $\Pi_{ir}$ in the edge. \Fref{fig:Er1}(c) shows the time-averaged $\Pi_{ir}$ versus $\psi_n$ at $t=0.8$ms. It is positive and increases towards the LCFS. Due to electrons' small mass, the total TAM flux comes predominantly from ions, $\Pi_{r}=\sum_s\Pi_{sr}\approx \Pi_{ir}$.  \Fref{fig:Er1}(d) compares the time-averaged $\pd_t\mc{L}_{\varphi}$ versus $-\pd_V\Pi_{r}$, and the two show reasonably close agreement. (Some numerical discrepancies are seen in the figure. These are due to flux-surface averaging and radial derivatives used in the data analysis; they do not affect the simulation itself.) This demonstrates the expected relation between the counter-current ion radial TAM flux, and the co-current toroidal rotation acceleration. 

Note that the total TAM $\mc{L}_\varphi=\sum_s\mc{L}_{s\varphi}+\mc{L}_E$ consists of the parallel-flow parts and $\ExB$ parts of ions and electrons. However, the electrons' parallel-flow TAM density is much smaller than ions' due to their small mass, $\mc{L}_{e\varphi}\ll\mc{L}_{i\varphi}$. The $\ExB$ TAM $\mc{L}_E$ is also much smaller than $\mc{L}_{i\varphi}$ (\fref{fig:Er1}(e)). Therefore, the negative $\pd_t\mc{L}_\varphi$ shown in \fref{fig:Er1}(d) mostly come from the parallel flow of ions. 

Finally, \fref{fig:Er1}(f) shows a comparison between the simulated $E_r$ and neoclassical theory outlined in \sref{sec:neo}. From \eref{Er_neo}, the neoclassical $E_r$ is different from $\pd_r p_i/en_i$ due to nonzero $\pd_r T_i$ and $\mc{L}_\varphi$. At $\psi_n<0.96$ where $\mc{L}_\varphi\approx 0$, the following fitting formula is found to match the simulated $E_r$: 
\begin{equation}
\label{Er_fit}
    E_r^{\rm fit}=\pd_r p_{i}/en_i-0.4\pd_r T_i/e.
\end{equation}
This formula does not match $E_r$ at $\psi_n>0.96$, possibly because it does not include the contribution from $\mc{L}_\varphi$. However, even if we use another fitting formula that includes $\mc{L}_\varphi$,
\begin{equation}
\label{Er_fit2}
    E_r^{\rm fit2}=\pd_r p_{i}/en_i-0.4\pd_r T_i/e-\frac{\mc{L}_{\varphi}RB_\theta}{m_in_iI^2\avg{B^{-2}}},
\end{equation}
the agreement is still not good. In particular, the finite $\mc{L}_\varphi$ gives a rather large correction to $E_r^{\rm fit}$, which suggests that the toroidal-rotation acceleration should significantly change $E_r$ according to the neoclassical theory. As seen from \fref{fig:Er1}(b), $\mc{L}_\varphi$ changes by about $\mathrm{10^{-4}kg/(m\cdot s)}$ from $t=0.8$ms to $t=1.2$ms. From \eref{Er_fit2}, $E_r$ should shift accordingly in the positive direction by about $2\times 10^3{\rm V/m}$. But $E_r$ from simulations does not change that much, further indicating that the neoclassical formula is not reproducing the observed $E_r$. To better understand the relation between toroidal rotation and $E_r$ in the edge, we take a closer look at the edge plasma profile in \sref{sec:flow} below.

As a side note, the neoclassical theory predicts $E_r=\pd_rp_i/en_i-k\pd_rT_i/e$ assuming zero TAM density, where $k=-2.1$ in the Pfirsch--Schl\"{u}ter regime, $k=-0.5$ in the plateau regime, and $k=1.17$ in the banana regime to the lowest order in the inverse aspect ratio $(r/R)$ \cite{Hazeltine74}. For our simulation, the edge plasma collision frequency is in the banana regime, $\nu_{i}qR/v_{ti}\ll (r/R)^{3/2}$, if we use the collision frequency $\nu_i\approx 0.6{\rm ms}^{-1}$, safety factor $q\approx 6$, ion thermal velocity $v_{ti}\approx 1.5\times 10^{5}{\rm m/s}$, minor radius $r\approx 0.6{\rm m}$, and major radius $R\approx 1.7{\rm m}$ for the calculation. The fitting formula \eref{Er_fit} corresponds to $k=0.4$, which falls in the range between the plateau-regime and the banana-regime limits.
\subsection{The balance of density flows in the edge}
\label{sec:flow}
\begin{figure*}
    \centering
    \includegraphics[width=0.65\textwidth]{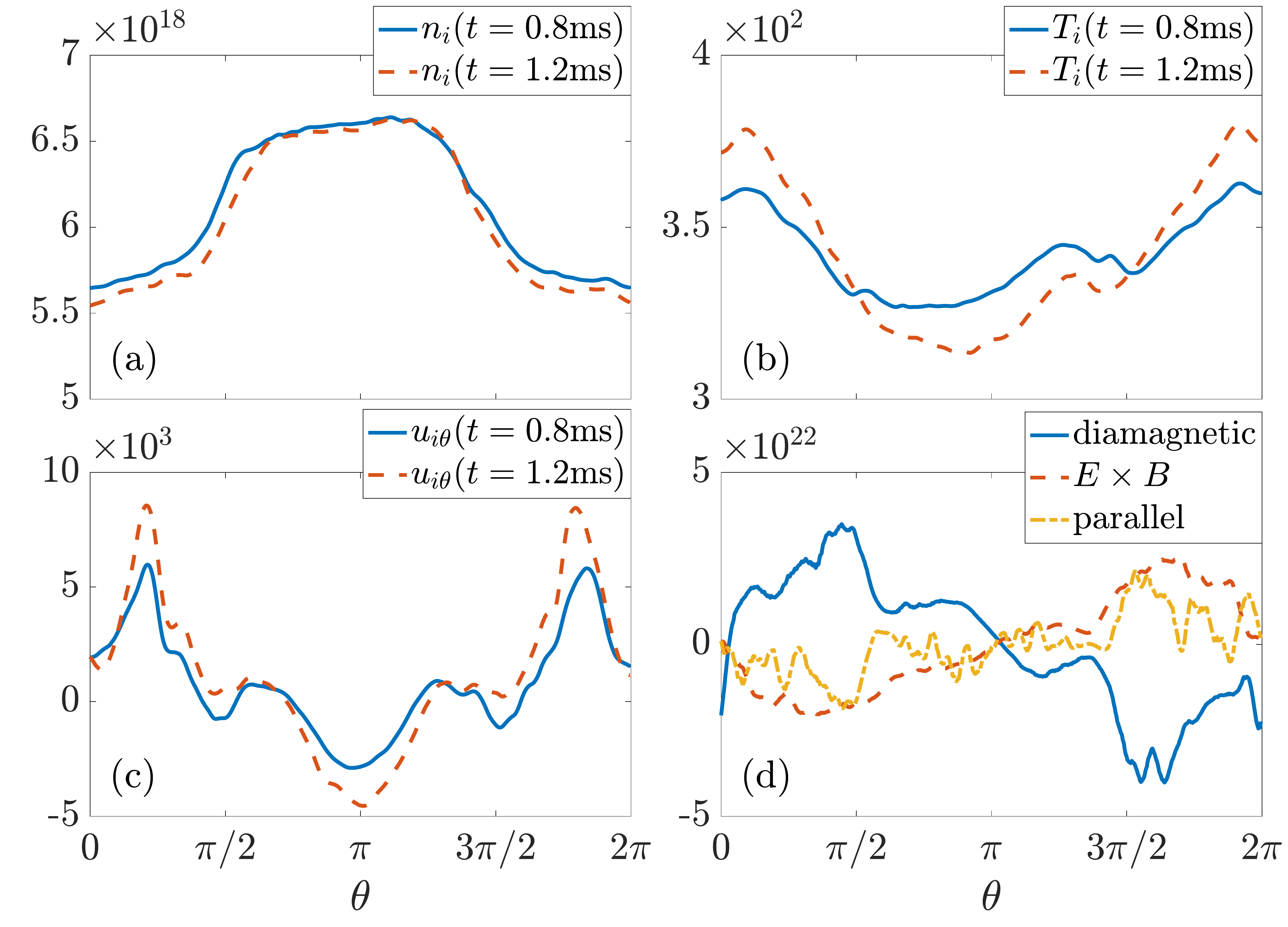}
    \caption{(a) The ion density (in units ${\rm m}^{-3}$) versus $\theta$ at the $\psi_n=0.99$ surface at $t=0.8$ms and $t=1.2$ms. (b) The ion temperature (in units eV). (c) The ion poloidal flow velocity \eref{utheta} (in units m/s). (d) The divergence of the density flow (in units ${\rm m^{-3}/s}$) from the diamagnetic flow, the $\ExB$ flow, and the parallel flow, separately. The data has been smoothed over the poloidal window $\Delta\theta=0.08\pi$.}
    \label{fig:flow1}
\end{figure*}

The plasma 2-dimensional profiles exhibit a significant poloidal asymmetry in the edge. Figures \ref{fig:flow1}(a) and (b) show the ion gyrocenter density $n_i=\int d\mc{W}F_i$ and temperature  $T_i=n_i^{-1}\int d\mc{W}(\mu B+p_\parallel^2/2m_i)F_i$ at two different moments of time.  The poloidal variation in $n_i$ and $T_i$ is about 20\%. This means that the neoclassical expression \eref{neo_flow} may not work as $n_i$ and $T_i$ are not flux functions, so we directly look at the ion fluid velocity given by \eref{fluid_flow}. \Fref{fig:flow1}(c) shows the poloidal flow 
\begin{equation}
\label{utheta}
u_{i\theta}=u_{i\parallel}B_\theta/B+E_rB_\varphi/B^2-(\pd_rp_i)B_\varphi/Z_ien_iB^2    
\end{equation}
versus the poloidal angle $\theta$, which is positive at the outboard ($\theta=0$) and negative at the inboard ($\theta=\pi$). This is in contrast to the neoclassical expression \eref{neo_flow}, where $u_{i\theta}=KB_\theta/n_i$ does not change sign. Therefore, even though the toroidal-rotation acceleration is co-current, $E_r$ does not necessarily shift in the positive direction as predicted by  neoclassical theory.

Despite the complicated density, temperature, and flow profile, the plasma remains nearly incompressible in the edge. \Fref{fig:flow1}(d) shows the divergence of the density flow $\nabla\cdot(n_i\bd{u}_i)$ from the diamagnetic term, the $\ExB$ term, and the parallel-flow term in \eref{fluid_flow}. The divergence of each component of the density flow is on the order of $10^{22}{\rm m^{-3}/s}$, but they roughly cancel each other when adding together. This is consistent with the slow time evolution of the density, which from \fref{fig:flow1}(a) can be estimated to be $\pd_t n_i\sim 10^{21}{\rm m^{-3}/s}$, much smaller than the divergence of each separate component of the density flows. Note that when calculating the divergence, the spatial derivatives can introduce significant numerical noise, which manifests as the high-harmonic (in $\theta$) fluctuations shown in \fref{fig:flow1}(d). This numerical noise only arises when post-processing the data using spatial derivatives, and does not affect the simulation itself.
\subsection{The ion gyrocenter particle and momentum flux}
\label{sec:results_flux}
\begin{figure*}
    \centering
    \includegraphics[width=0.65\textwidth]{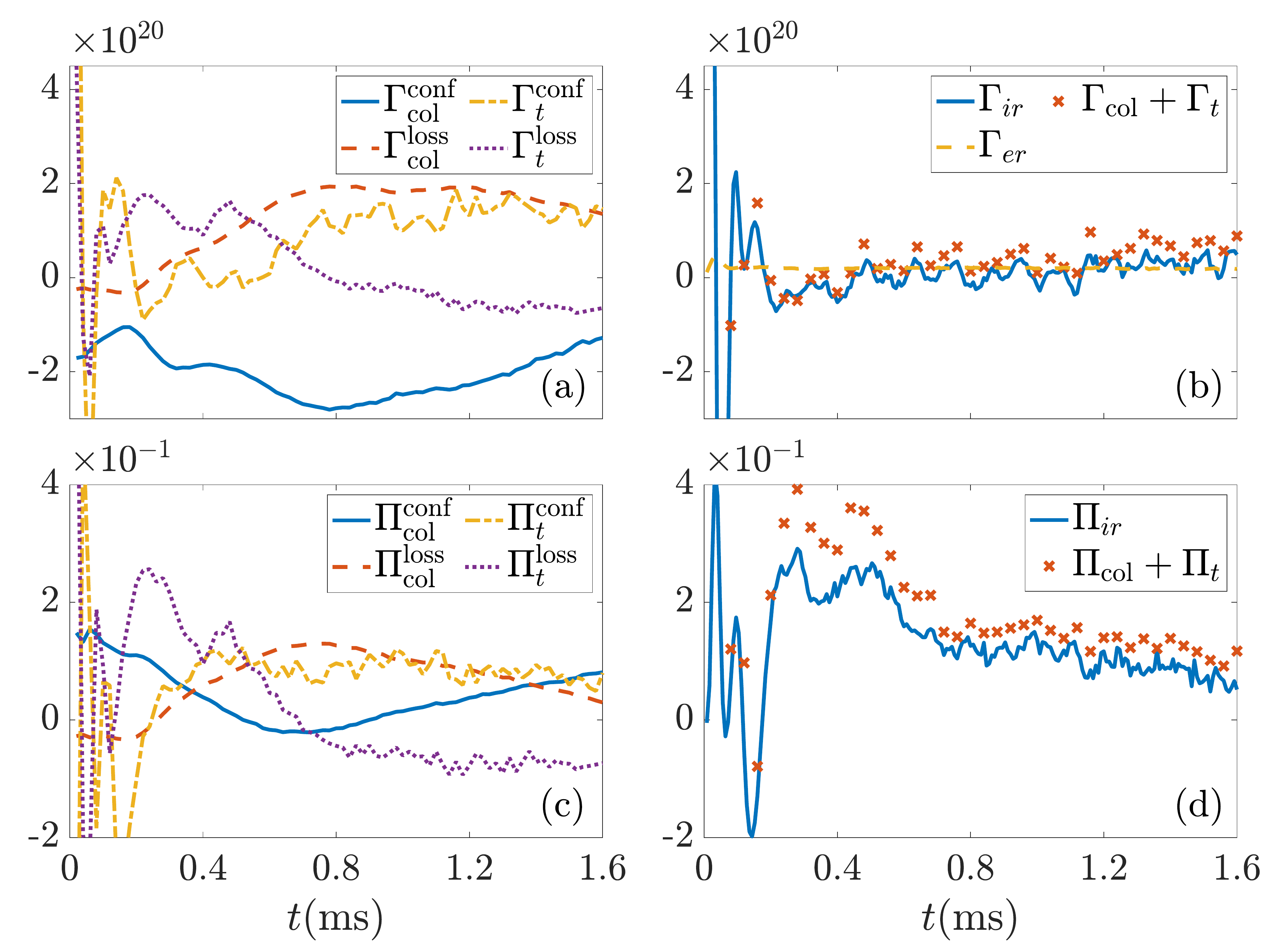}
    \caption{(a) The ion gyrocenter particle fluxes (in units ${\rm s}^{-1}$) across the $\psi_n=0.99$ flux surface. (b) Comparison between the orbit-flux results and XGC's diagnostic results. (c) The TAM flux (in units $\mathrm{kg\cdot m^2/s^2}$) of ion gyrocenters across the $\psi_n=0.99$ flux surface. (d) Comparison between the orbit-flux results and XGC's diagnostic results.}
    \label{fig:flux1}
\end{figure*}

The orbit-flux formulation (\sref{sec:orbit_flux}) has been used to calculate the ion radial gyrocenter particle and TAM flux across the $\psi_n=0.99$ flux surface. Since neutral ionization and heating happen mostly in the SOL, they do not directly contribute to orbit fluxes if we integrate along the part of the orbit inside this flux surface. Therefore, the ion radial gyrocenter flux can be written as the summation of four terms:
\begin{equation}
\label{flux_inside}
    \Gamma_{ir}=\Gamma_{\rm col}^{\rm conf}+\Gamma_{\rm col}^{\rm loss}+\Gamma_{t}^{\rm conf}+\Gamma_{t}^{\rm loss},
\end{equation}
which are shown in \fref{fig:flux1}(a). Within our simulation time span ($\Delta t\approx 1.6$ms), we have measured a positive collisional loss-orbit flux, $\Gamma_{\rm col}^{\rm loss}\approx 2\times 10^{20}{\rm s}^{-1}$, and a negative collisional confined-orbit flux, $\Gamma_{\rm col}^{\rm conf}\approx -2\times 10^{20}{\rm s}^{-1}$. This means that while collisions continuously scatter ions into the loss orbits, they also scatter ions out of the confined orbits. Note that the amplitude of both $\Gamma_{\rm col}^{\rm loss}$ and $\Gamma_{\rm col}^{\rm conf}$ decreases over time, indicating significant time evolution of the plasma. Such time evolution can also be seen from the nonzero $\Gamma_t$ terms. In particular, $\Gamma_t^{\rm loss}< 0$, meaning that the number of ions residing in the loss orbits increases over time. Meanwhile, $\Gamma_t^{\rm conf}>0$, so the number of ions residing in the confined orbits decreases over time. It is possible that as more ions are collisionally scattered into the loss orbits, the velocity-space gradient weakens at the boundary between loss orbits and confined orbits, causing the scattering rate (namely, $\Gamma_{\rm col}^{\rm loss}$) to decrease. Also, as mentioned in \sref{sec:setup2}, the ion temperature decreases over time in the confined region via collisions with electrons, which may also cause $\Gamma_{\rm col}^{\rm loss}$ to decrease.

\Fref{fig:flux1}(b) compares the orbit flux results $\Gamma_{\rm col}+\Gamma_t$ with the ion gyrocenter radial flux directly calculated by XGC using the distribution function,
\begin{equation}
    \Gamma_{ir}=\int d\bd{S}\cdot\int d\mc{W} F_i{\dot{\bd{R}}}.
\end{equation}
Note that $\Gamma_{ir}$ is calculated from integration over the flux surface, while $\Gamma_{\rm col}+\Gamma_t$ is calculated from integration over the orbits that cross the flux surface. The good agreement between the two demonstrates that our orbit-flux formulation is implemented with good numerical accuracy. Also shown in \fref{fig:flux1}(b) is the electron gyrocenter radial flux
\begin{equation}
    \Gamma_{er}=\int d\bd{S}\cdot\int d\mc{W} F_e{\dot{\bd{R}}}.
\end{equation}
It is seen that $\Gamma_{er}\approx 2\times 10^{19}{\rm s}^{-1}$ is nonzero, and $\Gamma_{ir}$ fluctuates around this value, too. Therefore, the radial gyrocenter flux is nonzero but ambipolar, and hence $E_r$ does not change much over time. 

Similar calculations are done for the ion gyrocenter radial TAM fluxes from \eref{formulation_flux2}:
\begin{equation}
\label{mflux_inside}
    \Pi_{ir}=\Pi_{\rm col}^{\rm conf}+\Pi_{\rm col}^{\rm loss}+\Pi_{t}^{\rm conf}+\Pi_{t}^{\rm loss}.
\end{equation}
These fluxes are shown in \fref{fig:flux1}(c), and a comparison with the direct TAM flux calculation by XGC,
\begin{equation}
    \Pi_{ir}=\int d\bd{S}\cdot\int d\mc{W} (\mc{P}_\varphi-Z_ie\psi)F_i{\dot{\bd{R}}},
\end{equation}
is shown in \fref{fig:flux1}(d). The collisional loss-orbit TAM flux is positive, $\Pi_{\rm col}^{\rm loss}\approx 0.1\mathrm{kg\cdot m^2/s^2}$. This is consistent with the theoretical expectation that loss orbits are mostly counter-current, thus ions on these orbits move counter-current TAM out of the flux surface. However, similar to $\Gamma_{\rm col}^{\rm loss}$, here $\Pi_{\rm col}^{\rm loss}$ also decreases over time.  Meanwhile, the collisional confined-orbit TAM flux $\Pi_{\rm col}^{\rm conf}$ is overall positive too and increases over time. Therefore, collisions cause both the loss orbits and the confined orbits to contribute to positive radial TAM fluxes. Also, just like the particle flux, here for the TAM flux the time-derivative terms are significant too. In particular, $\Pi_t^{\rm conf}>0$ and $\Pi_t^{\rm loss}<0$, meaning that the TAM carried by confined-orbit ions decreases (becomes more counter-current), while the TAM carried by loss-orbit ions increases (becomes more co-current).

We note that in the literature, ion orbit loss has sometimes been linked to counter-current toroidal rotation of the bulk plasma \cite{Helander05NBI,Thyagaraja07}, which differs from the co-current rotation observed in our simulations. The idea is simple: lost ions carry a radially outgoing current $\bd{J}_{\rm loss}$ and hence experience a Lorentz force $\bd{J}_{\rm loss}\times\bd{B}$ in the co-current direction.  Due to the quasineutrality constraint, the remaining bulk plasma must develop a radially incoming return current $\bd{J}_{\rm ret}=-\bd{J}_{\rm loss}$, and hence experience a Lorentz force $\bd{J}_{\rm ret}\times\bd{B}$ in the counter-current direction. Therefore, when looking at the remaining bulk plasma, the rotation acceleration would be counter-current. This effect is included in the gyrokinetic formulation: when ion gyrocenters are lost, the plasma develops a polarization return current to maintain quasineutrality, which corresponds to the left-hand side of the gyrokinetic Poisson equation \eref{XGC_poisson}. Consequently, $E_r$ shifts in the negative direction, and the corresponding $\mc{L}_E$ \eref{TAM_LE} increases,  which accounts for the counter-current acceleration of the bulk plasma.  However, this effect is not significant in our simulation:  while there are gyrocenter ions leaving the bulk plasma following the loss orbits, there are also gyrocenter ions entering the bulk plasma at a similar rate following the confined orbits (\fref{fig:flux1}(a)).  Therefore, the return current is small and $E_r$ does not change much at $t>0.8$ms.

We also note that although effects from neutral dynamics and heating do not show up explicitly in the above orbit-flux results, they still have significant impact on the simulation results. As discussed in \ref{sec:SOL_flux}, one can  evaluate their contribution to the confined-orbit radial fluxes if the integration is along the orbits outside the flux surface.

In summary, for the gyrocenter particle flux, the loss-orbit contribution is nearly balanced by the confined-orbit contribution, leaving a small positive ambipolar value in the sum. However, for the gyrocenter TAM flux, both the loss orbits and the confined orbits contribute to radially outward counter-current momentum flux. Therefore, while $E_r$ is quasisteady, the toroidal rotation still shifts in the negative (co-current) direction. Admittedly, these orbit-loss effects weaken over time using our current simulation setup, and presumbably would not last longer than the collisional time scale.
\subsection{The effects of collisional loss-orbit flux on radial electric field and toroidal rotation}
\begin{figure*}
    \centering
    \includegraphics[width=0.65\textwidth]{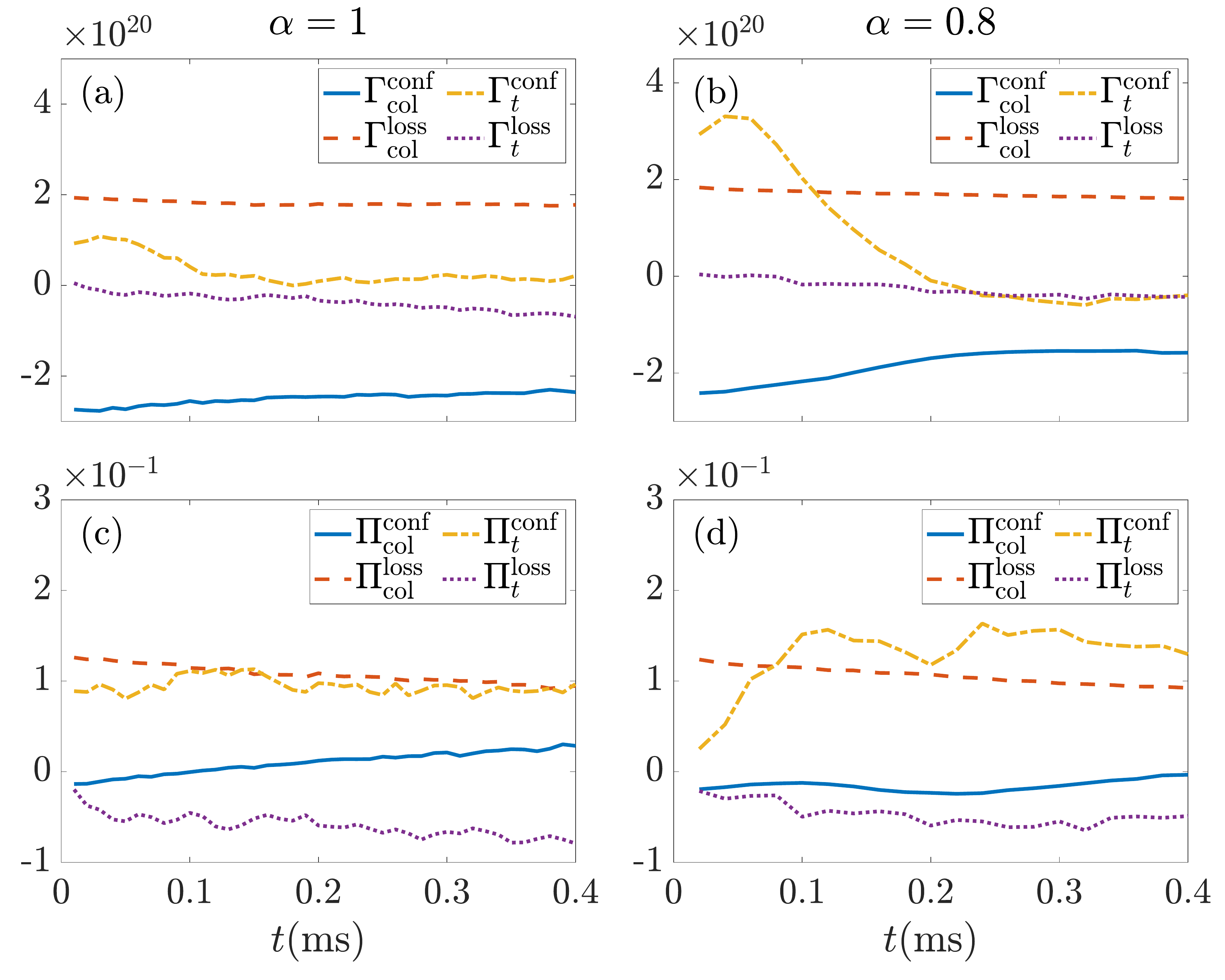}
    \caption{The ion gyrocenter particle and TAM fluxes across the $\psi_n=0.99$ flux surface in fixed-$E_r$ simulations, where $E_r$ is given by \eref{restart_Er}. Results with $\alpha=1$ are shown in (a) and (c). Results with $\alpha=0.8$ are shown in (b) and (d).}
    \label{fig:flux_restart}
\end{figure*}

Since the positive loss-orbit gyrocenter flux is balanced by the negative confined-orbit gyrocenter flux, the steady-state $E_r$ is expected to be different from that in a plasma without orbit loss, where the confined-orbit flux alone should vanish. However, loss orbits are an integral part of the edge plasma, so the loss-orbit effects cannot be turned off. Therefore, we  evaluate how the loss-orbit flux shifts $E_r$ in the following way. We start a new simulation, in which the initial plasma distribution function is taken from the original simulation at $t=0.8$ms, when $E_r$ is already quasisteady. In contrast to the original simulation, where $E_r$ in the confined region is solved self-consistently using \eref{XGC_poisson}, in the new simulation we require the confined-region $E_r$ to be multiplied by a factor $\alpha$,
\begin{equation}
\label{restart_Er}
    E_r^{\rm new}=\alpha E_r^{\rm original}|_{t=0.8{\rm ms}},
\end{equation}
and fixed in time. (The positive SOL $E_r$ is not changed, since it is determined by different physics, unaffected by orbit loss.) We then measure the ion radial fluxes under this imposed fixed $E_r$.

Note that the ion gyrocenter orbits change due to the change in $E_r$. But the change only occurs inside the LCFS where $E_r$ is multiplied by $\alpha$. In the SOL, $E_r$ (and the orbits) remain the same. Therefore, the boundary between loss orbits and confined orbits does not change much in the $(\mu,\mc{P}_\varphi,H)$ space with two different values of $\alpha$, so that we can make direct comparisons over orbit fluxes. 

We chose two different values of $\alpha$: $\alpha=1$ and $\alpha=0.8$. The resulting ion gyrocenter particle fluxes across the $\psi_n=0.99$ surface are plotted in figures \ref{fig:flux_restart}(a) and (b), and the ion gyrocenter TAM fluxes are plotted in figures \ref{fig:flux_restart}(c) and (d). For $\alpha=1$ (figures \ref{fig:flux_restart}(a) and (c)), the fixed $E_r$ in the new simulation is similar to the self-consistent $E_r$ in the original simulation, although the latter slowly evolves in time. As shown in the figures, there are no abrupt changes in orbit fluxes, and the total flux remains small. For $\alpha=0.8$ (figures \ref{fig:flux_restart}(b) and (d)), the fixed $E_r$ is now very different from the self-consistent $E_r$. As shown in the figures, the loss-orbit fluxes remain similar, but the confined-orbit fluxes change significantly. In particular, for the confined-orbit gyrocenter flux, $\Gamma_t^{\rm conf}$ reaches a large positive value shortly after the beginning of the new simulation. This means that due to reduced $|E_r|$, there are more ions leaving than entering the flux surface, which results in the decrease in number of ions residing in confined orbits. Also, $\Gamma_{\rm col}^{\rm conf}$ shifts in the positive direction by about $10^{20}{\rm s}^{-1}$ within $0.2{\rm ms}$ compared to the case with $\alpha=1$. This is consistent with the general expectation that if $|E_r|$ is smaller than the self-consistent value, collisions will adjust accordingly and create a positive ion radial gyrocenter flux, such that $|E_r|$ gets larger.

We now estimate how much the loss-orbit flux can shift $E_r$. As shown in figures~\ref{fig:flux_restart}(a) and (b), shortly after the beginning of the simulations, the confined-orbit flux $\Gamma_t^{\rm conf}+\Gamma_{\rm col}^{\rm conf}$ is about $-1.5\times 10^{20}{\rm s}^{-1}$ for $\alpha=1$ and $1\times 10^{20}{\rm s}^{-1}$ for $\alpha=0.8$. From linear interpolation, it can thus be estimated that with a 12\% reduction in $E_r$, the confined-orbit flux alone will vanish. Namely, the loss-orbit flux can shift $E_r$ in the negative direction by 12\%, compared to that in a plasma without orbit loss, where the confined-orbit flux alone should vanish. We can also estimate how fast $\Gamma^{\rm loss}$ can shift $E_r$, if the latter is solved self-consistently in the confined region. From the gyrokinetic Possion equation \eref{XGC_poisson}, we have 
\begin{equation}
\label{gauss}
\int (\delta n_i-\delta n_e)\,dV=\int \frac{n_{i0}m_i}{eB^2} E_r\,dS\approx 2\times 10^{16}
\end{equation}
at the $\psi_n=0.99$ flux surface, assuming $\delta\bar{n}_i\approx\delta n_i$. Therefore, with a positive $\Gamma^{\rm loss}\approx 2\times 10^{20}{\rm s}^{-1}$, it takes $0.012{\rm ms}$ for $E_r$ to shift in the negative direction by 12\%, which is much shorter compared to the simulation time scale. In other words, if $|E_r|$ were reduced by 12\% and were allowed to evolve self-consistently as in the original simulation, the loss-orbit flux will restore its value almost instantaneously. However, since $E_r$ is fixed in the new simulation presented in this subsection, the confined-orbit flux adjusted instead of $E_r$. As seen in \fref{fig:flux_restart}(b), $\Gamma_t^{\rm conf}$ drops to a slightly negative value at later time, such that $\Gamma^{\rm conf}$ is again negative and balances $\Gamma^{\rm loss}$. 
\section{Conclusions}
\label{sec:conclusions}
We used XGCa to study collisional ion orbit loss in an axisymmetric DIII-D L-mode plasma with gyrokinetic ions and drift-kinetic electrons. Numerical simulations, in which the plasma density and temperature profiles are maintained through neutral ionization and heating, show the formation of a quasisteady negative $E_r$ in the edge. We have measured a radially outgoing ion gyrocenter flux due to collisional scattering of ions into the loss orbits, which is balanced by the radially incoming ion gyrocenter flux from confined orbits on the collisional time scale. This suggests that collisional ion orbit loss can shift $E_r$  in the negative direction, compared to plasmas without orbit loss. It is also found that collisional ion orbit loss can contribute to a counter-current TAM flux, which is not balanced by the TAM flux carried by ions on the confined orbits. Therefore, the edge toroidal rotation shifts in the co-current direction on the collisional time scale.
\ack
This work was supported by the U.S. Department of Energy under contract number DE-AC02-09CH11466. The United States Government retains a non-exclusive, paid-up, irrevocable, world-wide license to publish or reproduce the published form of this manuscript, or allow others to do so, for United States Government purposes. Funding to R. Hager, S. Ku and C.S. Chang is provided via the SciDAC-4 program. The simulations presented in this article were performed on computational resources managed and supported by Princeton Research Computing, a consortium of groups including the Princeton Institute for Computational Science and Engineering (PICSciE) and the Office of Information Technology's High Performance Computing Center and Visualization Laboratory at Princeton University.  This research used resources of the National Energy Research Scientific Computing Center, which is supported by the Office of Science of the U.S. Department of Energy under Contract No. DE-AC02-05CH11231.
\section*{Data Availability}
Digital data can be found in DataSpace of Princeton University \cite{data}.
\appendix
\section{Effects of neutral dynamics and heating on the confined-orbit fluxes}
\label{sec:SOL_flux}
\begin{figure*}
    \centering
    \includegraphics[width=0.65\textwidth]{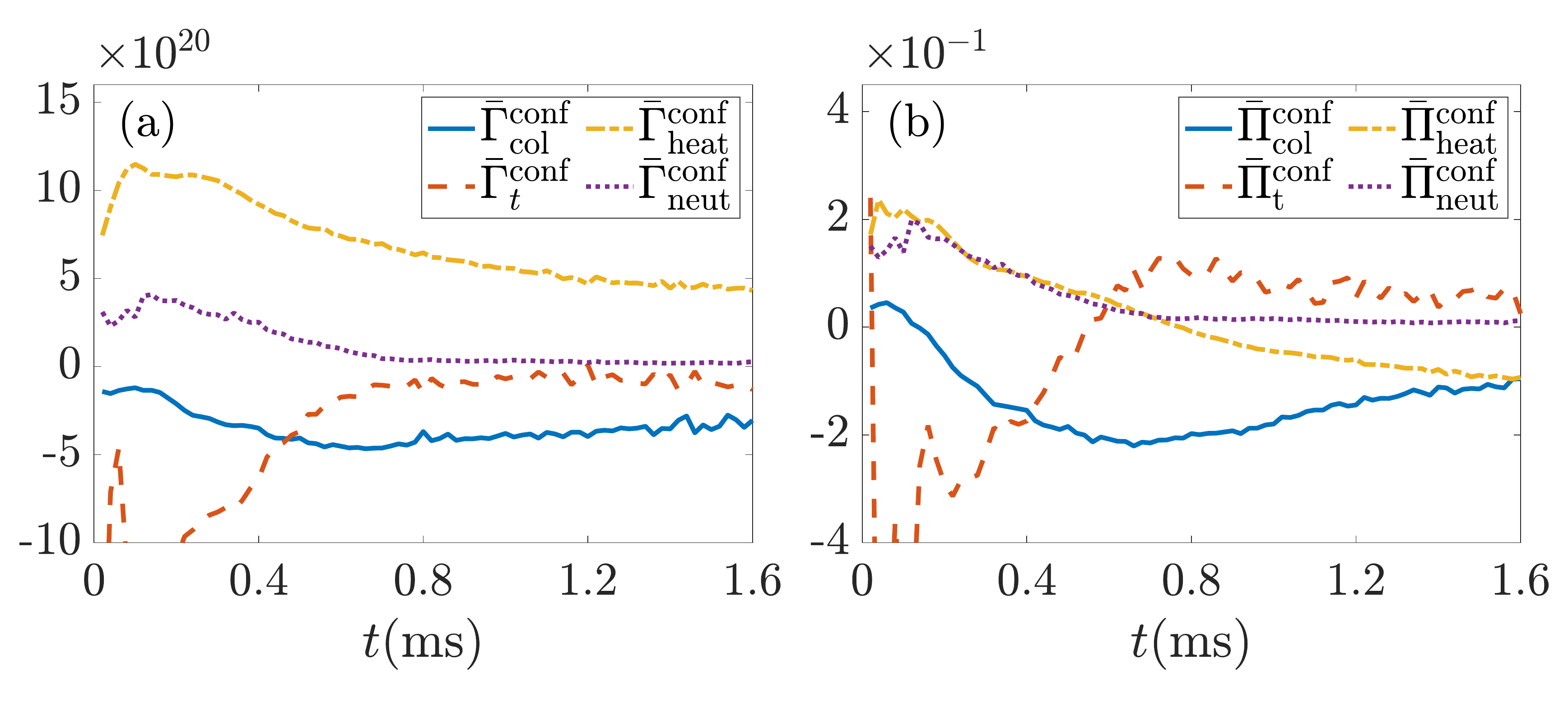}
    \caption{Confined-orbit gyrocenter particle fluxes (a) and TAM fluxes (b) calculated from integration outside the $\psi_n=0.99$ flux surface. Since we view the fluxes from outside the  surface, positive means radially inwards here.}
    \label{fig:flux1_sol}
\end{figure*}
In \sref{sec:results_flux}, orbit fluxes are calculated from orbit integration inside the $\psi_n=0.99$ surface, so the contributions from neutrals and heating do not explicitly show up. However, as an integrated part of the simulation, the plasma dynamics outside this surface, including the SOL, are also important for the simulation results. Since ions on loss orbits do not return to the confined region, whatever happens along outgoing loss orbits outside the flux surface will not be directly relevant. However, ions on confined orbits will later return and affect ion radial fluxes. Therefore, scattering of ions into and out of the confined orbits outside the $\psi_n=0.99$ flux surface will affect ion radial fluxes across this surface. These effects are included in the confined-orbit flux, and can be calculated if we integrate along the portion of the orbit that lays outside the $\psi_n=0.99$ surface.  In this way, we can also assess effects from neutral ionization and heating on ion radial fluxes. Following \eref{formulation_outside}, we expect \begin{equation}
\Gamma_{\rm col}^{\rm conf}+\Gamma_t^{\rm conf}=-(\bar{\Gamma}_{\rm col}^{\rm conf}+\bar{\Gamma}_{t}^{\rm conf}+\bar{\Gamma}_{\rm heat}^{\rm conf}+\bar{\Gamma}_{\rm neut}^{\rm conf}).
\end{equation}
Here, we put a bar on the right-hand-side $\Gamma$ terms to indicate that they are calculated from the integration outside the flux surface. Similarly, for the TAM fluxes, we have
\begin{equation}
\Pi_{\rm col}^{\rm conf}+\Pi_t^{\rm conf}=-(\bar{\Pi}_{\rm col}^{\rm conf}+\bar{\Pi}_{t}^{\rm conf}+\bar{\Pi}_{\rm heat}^{\rm conf}+\bar{\Pi}_{\rm neut}^{\rm conf}).
\end{equation}
These fluxes are plotted in \fref{fig:flux1_sol}. Note that since we view the fluxes from outside the surface, a positive flux means that the corresponding quantity is carried radially inwards. For the gyrocenter particle fluxes, the dominant contributions are from a negative $\bar{\Gamma}_{\rm col}^{\rm conf}$ and a positive $\bar{\Gamma}_{\rm heat}^{\rm conf}$. Therefore, outside the flux surface, collisions scatter ions off confined orbits, while heating put ions into the confined orbits. The gyrocenter TAM fluxes are also dominated by the collision term and the heating term, which are both negative at later time ($t>0.8{\rm ms}$). Effects from neutral ionization and charge exchange are comparable with collisions and heating at the beginning of the simulation, but become small at later times. These results are consistent with the confined-orbit fluxes shown in \fref{fig:flux1} and provide an alternative view of them from outside the flux surface.
\section{Dependence of collision ion orbit loss on the plasma density}
\label{sec:density}
\begin{figure*}
    \centering
    \includegraphics[width=0.65\textwidth]{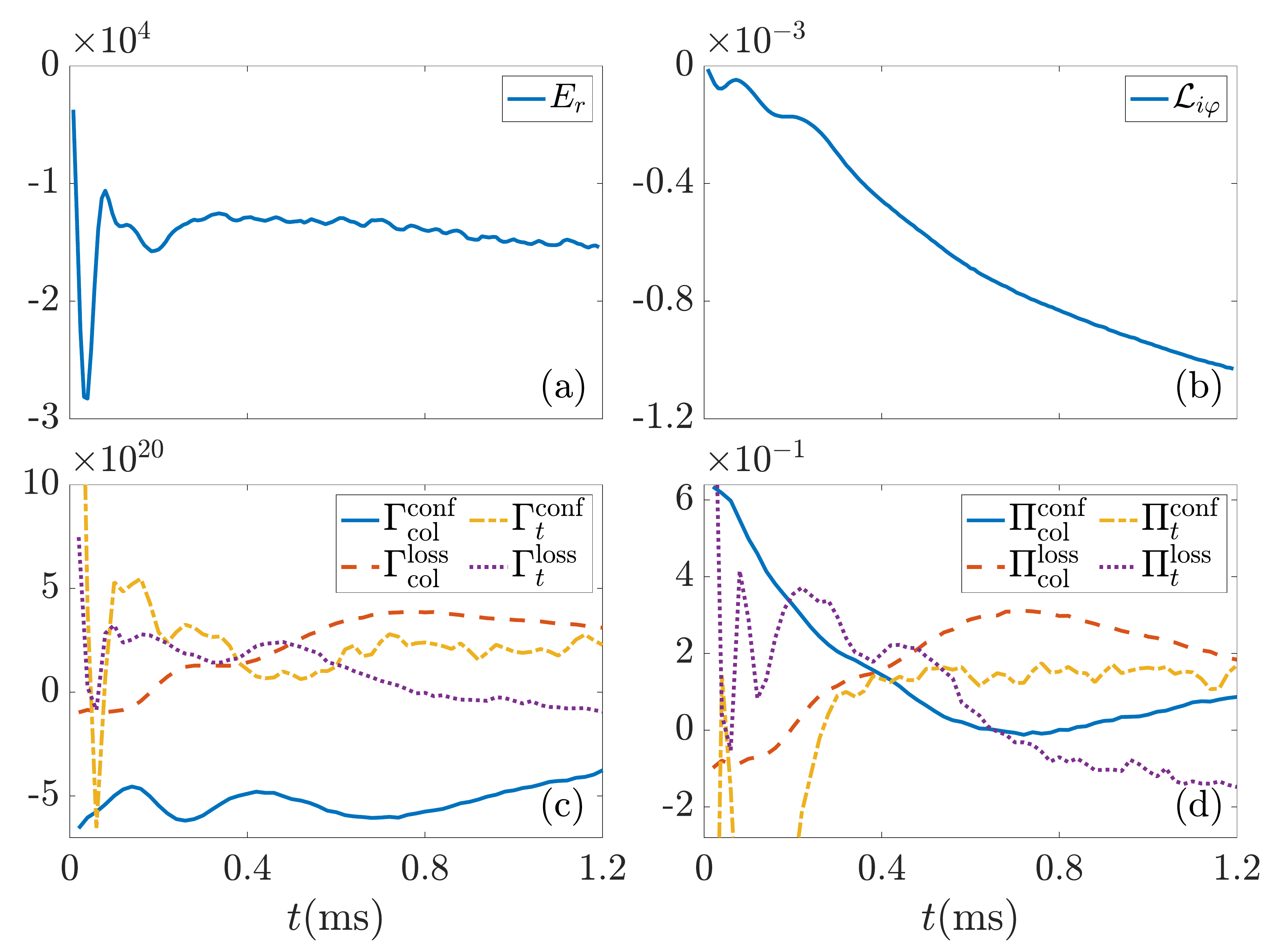}
    \caption{Simulation results at the $\psi_n=0.99$ flux surface with plasma density doubled.}
    \label{fig:flux_den2}
\end{figure*}
Naturally, one is interested in the dependence of collisional orbit-loss effect on the ion--ion collisionality, which is determined by $n_i$ and $T_i$. From neoclassical theory outlined in \sref{sec:neo}, $E_r$ is determined by $n_i^{-1}\pd_r(n_i T_i)$ and $\pd_r T_i$ assuming $\mc{L}_\varphi\approx 0$. Therefore, it is easier to first study the scaling dependence on $n_i$ instead of $T_i$, since $E_r$ and the phase-space structure of orbits do not change directly when $n_i$ is multiplied by a constant factor. Here, we present results on collisional ion orbit loss when the ion and electron density are doubled compared to that shown in \fref{fig:Lprofile}. The collision frequency $\nu_i$ is doubled accordingly, but the edge plasma is still in the banana regime, $\nu_i qR/v_{ti}\ll(r/R)^{3/2}$. Therefore, the neoclassical estimate of $E_r$ \eref{Er_neo} does not change when $n_i$ is doubled, and the phase-space boundary between confined orbits and loss orbits should remain similar.

The results of $E_r$ and toroidal rotation at $\psi_n=0.99$ are shown in figures \ref{fig:flux_den2}(a) and \ref{fig:flux_den2}(b). Similar to the results shown in \fref{fig:Er1}, here the toroidal-rotation acceleration is co-current while $E_r$ is quasisteady. Note that $E_r$ is similar to that in \fref{fig:Er1}(a), consistent with the neoclassical estimate. The TAM density is doubled compared to that in \fref{fig:Er1}(b) when $n_i$ is also doubled, so the corresponding fluid velocity remains similar.

The orbit-flux results are shown in figures \ref{fig:flux_den2}(c) and \ref{fig:flux_den2}(d), which are roughly doubled compared to those shown in \fref{fig:flux1}. This is different from the intuition that collisional orbit-loss effect is proportional to $\nu_i n_i$, which scales as $n_i^2$. (An earlier analytic study estimated this effect to be proportional to $n_i^{7/8}$ in the banana regime \cite{Shaing92b}.) It is possible that as collisionality goes up, the velocity-space gradient weakens at the boundary between loss orbits and confined orbits, causing the local scattering rate to decrease.
\section{Collisional ion orbit loss with reversed toroidal magnetic field}
For the simulation results presented in the main text, the ion grad-$B$ and curvature drift point in the negative-$z$ direction towards the X point. This is often referred  ``favorable'' configuration as opposed to the ``unfavorable'' configuration where ions drift away from the X point. Here, we briefly present simulation results with the unfavorable configuration. For the simulation setup, the sign of $B_\varphi$ is reversed such that the ion grad-$B$ and curvature drift reverse their sign in the $z$ direction, but all other settings remain the same.

\begin{figure*}
    \centering
    \includegraphics[width=0.65\textwidth]{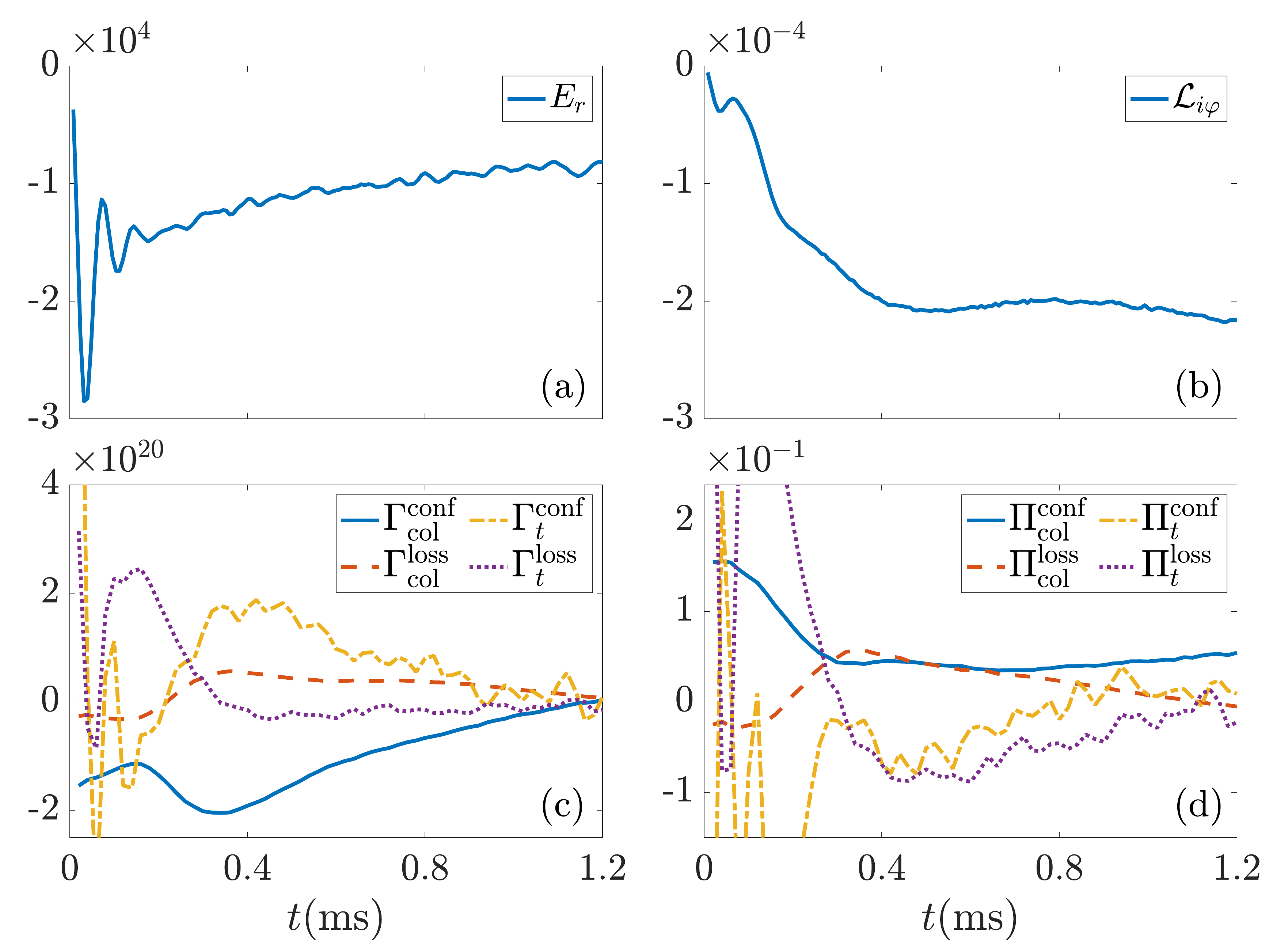}
    \caption{Simulation results at the $\psi_n=0.99$ flux surface with $B_\varphi$ reversed.}
    \label{fig:backward}
\end{figure*}

The results are shown in \fref{fig:backward} for the $\psi_n=0.99$ flux surface. Compared to those in \fref{fig:Er1}, $E_r$ is much weaker in this case. However, note that $n_i$ and $T_i$ are not maintained exactly within our simulation setup, and their profiles are also different in this case at later time of the simulation, which could cause the difference in $E_r$. A more notable difference lies in the toroidal rotation, which is less sensitive to the $n_i$ and $T_i$ profiles. As shown in the figure, the co-current toroidal-rotation acceleration ends very early. These results suggest that collisional ion orbit loss is weaker in this case, which is indeed corroborated by the orbit-flux results. Both the collisional loss-orbit gyrocenter particle flux $\Gamma_{\rm col}^{\rm loss}\approx 5\times 10^{19}{\rm s}^{-1}$ and TAM flux $\Pi_{\rm col}^{\rm loss}\approx 5\times 10^{-2}\mathrm{kg\cdot m^2/s^2}$ are smaller compared to those shown in \fref{fig:flux1}. Also, for the TAM flux, the collisional terms $\Pi_{\rm col}^{\rm conf}$ and $\Pi_{\rm col}^{\rm loss}$ are almost balanced by the transient terms $\Pi_t^{\rm conf}$ and $\Pi_t^{\rm loss}$, so the total TAM flux is very small.

Therefore, for the unfavorable configuration, we observe reduced collisional ion orbit loss associated with weakened $E_r$ and toroidal rotation. It is straightforward to show that for an axisymmetric $H$, the ion's equation of motion is invariant under $(B_\varphi,p_\parallel,t)\to(-B_\varphi,-p_\parallel,-t)$. Therefore, under similar level of $E_r$, there is no significant difference in ion orbits between the favorable and unfavorable configurations, and the observed difference in collisional ion orbit loss must be due to something else. It is often argued that loss-orbit ions travel longer distances from the LCFS to the wall for the unfavorable configuration, so that they are more likely scattered back into confined region via collisions. However, such effect should result in increased $F_i^{\rm in}$ for the confined orbits, and hence increased magnitude of $\Gamma^{\rm conf}$ (assuming it is negative). More study is thus required to identify what directly causes the reduced $\Gamma^{\rm loss}$ and $\Pi^{\rm loss}$.

\label{sec:backward}
\section*{References}
\bibliographystyle{iopart-num}
\bibliography{references}
\end{document}